\documentclass[epj,final,amsmath,amssymb,pdftex]{svjour}
\pdfoutput=1

\def\<{{<}}
\def\>{{>}}

\usepackage{graphicx}
\usepackage{amssymb}

\input{epsf}

\begin{document}

\title{Ulam method for the Chirikov standard map}

\author{Klaus M. Frahm and Dima L. Shepelyansky}

\institute{Laboratoire de Physique Th\'eorique (IRSAMC), 
Universit\'e de Toulouse, UPS, F-31062 Toulouse, France
\and
LPT (IRSAMC), CNRS, F-31062 Toulouse, France
\and
http://www.quantware.ups-tlse.fr}

\titlerunning{Ulam method for the Chirikov standard map}
\authorrunning{K.M.Frahm and D.L.Shepelyansky}

\date{submitted: April 8, 2010, accepted for EPJ B: May 20, 2010 }
%\date{\today}

\abstract{We introduce a generalized Ulam method and
apply it to symplectic dynamical maps with a divided phase space.
Our extensive numerical studies 
based on the Arnoldi method show that the Ulam approximant of the
Perron-Frobenius operator on a chaotic component converges to
a continuous limit. Typically, in this regime the spectrum
of relaxation modes is characterized by a power law
decay for small relaxation rates. Our numerical data show that the exponent
of this decay is approximately equal 
to the exponent of Poincar\'e recurrences
in such systems. The eigenmodes show links with 
trajectories sticking around stability islands.
}

\maketitle
%%%%%%%%%%%%%%%%%%%%%%%%%%%%%%%%%%%%%%%%%%%%%%%%%%%%%%%%%
\section{Introduction}
\label{sec1}

The properties of two-dimensional (2D) symplectic maps
with dynamical chaos have been studied in great detail during
last decades both on mathematical 
(see e.g. \cite{arnold,sinai} and Refs. therein) and physical
(see e.g. \cite{chirikov1969,chirikov1979,lichtenberg} 
and Refs. therein)
grounds. A generic and  nontrivial behavior appears in maps with
divided phase space where islands of stability
are surrounded by chaotic  components. A typical example of such 
a map is the Chirikov standard map \cite{chirikov1969,chirikov1979}
which often gives a local description of dynamical chaos
in other dynamical maps and describes
a variety of physical systems (see e.g. \cite{scholar}).
This map is characterized by one dimensionless chaos parameter $K$
and two dynamical variables $x,y$ which have a meaning of phase and
conjugated action:
\begin{equation}
\label{eq_stmap}
{\bar y} = y + \frac{K}{2\pi} \sin (2\pi x) \; , \;\; 
{\bar x} = x + {\bar y} \;\; ({\rm mod} \; 1) \;.
\end{equation}
Here bars mark the variables after one map iteration
and we consider the dynamics to be periodic on  a torus so that
$0 \leq x \leq 1$, $0 \leq y \leq 1$.

For small values of $K$ the phase space  is covered by invariant
Kolmogorov-Arnold-Moser (KAM) curves which restrict
dynamics in action variable $y$.
With the increase of $K$ more and more of these KAM curves start
to be destroyed and above a certain $K_c$ all curves
disappear and dynamics in $y$  becomes unbounded.
In 1979  Greene \cite{greene}
argued that the last KAM curve has the golden rotation
number $r=r_g=\langle(x_t-x_0)/t\rangle=(\sqrt{5}-1)/2$ with the critical
$K_g =0.9716...$ (here $t$ is given in number of map iterations;
there is also symmetric critical curve at $r=1-r_g$ at $K_g$).
A renormalization technique
developed by MacKay \cite{mackay} allowed 
to determine $K_g=0.971635406$
with enormous precision. The properties of the critical golden curve
on small scales are universal for all critical curves with the 
golden tail of the continuous fraction expansion of $r$
for all smooth 2D symplectic maps \cite{mackay}.
Further mathematical \cite{percival}
and numerical \cite{chirikov2000}
results showed that the actual value of $K_c$ 
is indeed very close to $K_g$
($K_c-K_g < 2.5 \times 10^{-4}$ according to \cite{chirikov2000})
and it is most probable that $K_c=K_g$.

The results of Greene and MacKay \cite{greene,mackay}
gave a fundamental understanding of the local structure properties of 
symplectic maps in a vicinity of critical invariant curves
but the global properties of 
dynamics on a chaotic component  still keep their mysteries.
For $K>K_g$ the golden KAM curve is replaced 
by a cantori \cite{aubry}
which can significantly affect the diffusive transport
through the chaotic part of the phase space \cite{meiss,stark}.
In addition there are other internal boundaries of the chaotic
component with critical
invariant curves which can affect statistical properties 
of chaotic dynamics. One of such important
properties is the statistics of Poincar\'e recurrences
$P(t)$ which is characterized by a slow algebraic decay in time 
being in contrast to an exponential decay in a homogeneously 
fully chaotic maps
(see \cite{kiev,karney,chsh,ott,chirikov1999,ketzmerick,artuso}
and Refs. therein). This algebraic decay
$P(t) \propto 1/t^\beta$ has $\beta \approx 1.5$. 
Such a slow decay appears due to trajectory
sticking near stability islands and critical invariant 
curves and leads to even slower correlation decay
with a divergence of certain second moments. 
A detailed understanding of this phenomenon 
is related to global properties of dynamical chaos
in 2D symplectic maps and is still missing.

With the aim to analyze the global properties
of chaotic dynamics we use the Ulam method 
proposed in 1960 \cite{ulam}. In the original version of 
this method the phase space is divided in
$N_d=M \times M$ cells and $n_c$ trajectories
are propagated on one map iteration from each cell $j$.
Then the matrix $S_{ij}$ is defined by the relation
$S_{ij}= n_{ij}/n_c$ where $n_{ij}$ is the number of trajectories
arrived from a cell $j$ to a cell $i$.
By the construction $\sum_i S_{ij}=1$ and hence the
matrix $S_{ij}$ belongs to the class of the Perron-Frobenius
operators (see e.g. \cite{mbrin}) and can be considered
as a discrete Ulam approximate of the Perron-Frobenius
operator (UPFO) of the continuous dynamics. 
According to the Ulam conjecture \cite{ulam}
the UPFO converges to the continuous limit
at large $M$. Indeed, this conjecture was proven
for one-dimensional (1D) homogeneously chaotic
maps \cite{li}. Various properties of the UPFO
for 1D maps have been studied in 
\cite{tel,kaufmann,froyland2007} and further mathematical
results have been reported in 
\cite{ding,liverani,froyland2008a,froyland2008b}
with extensions to 2D maps. 
It was also shown that the UPFO can find useful applications in studies of 
dynamics of molecular systems
\cite{schutte} and  coherent structures in dynamical flows
\cite{froyland2009physd}.
Recent studies 
\cite{zhirov,ermann} traced similarities 
between the UPFO, the corresponding  to them Ulam networks 
and the properties of the Google matrix of the world wide web
networks.

While for homogeneously chaotic systems
the Ulam method is well convergent to a
continuous limit it is also well known that in certain cases
the discretization leads to violent modifications of
system properties
(see e.g. \cite{liverani}). 
For example, for 2D maps with a divided phase space
the UPFO destroys all KAM curves and thus  
absolutely modifies the system properties (see e.g. discussion
in \cite{zhirov}). The physical origin of these
unacceptable modifications is related to a small
noise, introduced by the coarse-graining,
which amplitude is proportional to the cell size $1/M$.
This noise allows  trajectories to penetrate through
invariant curves leading to a broadly 
known opinion that the Ulam method is not 
applicable to the Hamiltonian systems
with divided phase space.

In this work we show that the Ulam method can be generalized
in such a way that it becomes applicable to 2D symplectic maps
with a divided phase space. We use this generalized
Ulam method to investigation of the Chirikov standard
map at the critical parameter $K_g$
and at large values of $K$ when the phase space
has small stability islands. Our extensive 
numerical simulations allow to obtain
new features of the global chaotic dynamics in such 
cases. We also show that this method 
can be applied to other maps, e.g. the separatrix map
or whisker map \cite{chirikov1979}.

The paper is constructed as follows: in Section 2
we describe the generalized Ulam method 
and demonstrate its convergence for the map (\ref{eq_stmap})
at $K=K_g$, in Section 3 we describe
the Arnoldi method which allows to study the spectral
properties of the UPFO in the limit of large matrix
size up to $N_d \sim 10^6$. The spectral
properties of the UPFO 
are analyzed in Section 4 for the map (\ref{eq_stmap}) at $K=K_g$
and in Section 5 at $K=7$. The case of the 
separatrix map with the critical golden curve
is studied in Section 6, the discussion of the results is
presented in Section 7.

%%%%%%%%%%%%%%%%%%%%%%%%%%%%%%%%%%%%%%%%%%%%%%%%%%%%%%%%%
\section{Generalized Ulam method}

\begin{figure}[h]
\begin{center}
\includegraphics[width=0.48\textwidth]{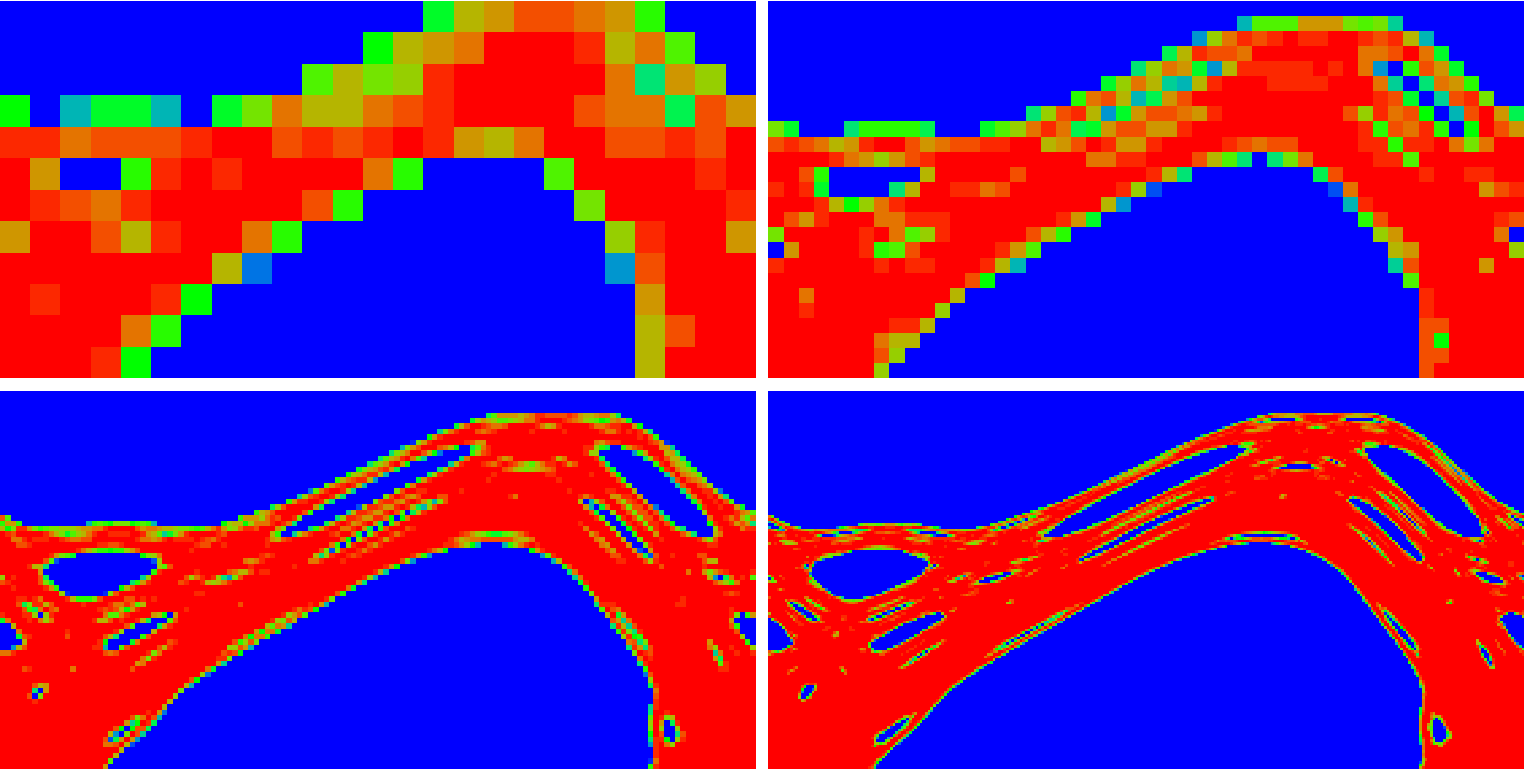}
\end{center}
\caption{\label{fig1} (Color online)
Density plots of the eigenvector $\psi_0$ of the UPFO 
with eigenvalue $\lambda_0=1$. The UPFO is obtained by the 
generalized Ulam method with a single trajectory of 
$10^{12}$ iterations of the Chirikov standard map (\ref{eq_stmap}) 
at $K=K_g=0.971635406$. 
The phase space is shown in the area
$0\leq x \leq 1$, $0 \leq y \leq 1/2$; 
the UPFO is obtained from $M \times M/2$ cells
placed in this area.
The value of $M$ for the panels is $25$ (first/top left),
$50$ (first/top right), $140$ (second/bottom left),
$280$ (second/bottom right), the corresponding dimension of the
UPFO matrix $S$ is $N_d=177, 641, 4417, 16609$
respectively. The probability density of the eigenstate
is shown by color with red/grey for maximum and blue/black for
zero.
}
\end{figure}

To make the Ulam method to be applicable for the symplectic maps
with divided phase space we use the following generalization 
of the method which we explain on an example of 
the Chirikov standard map (\ref{eq_stmap}). The whole phase
space $0\leq x \leq 1$, $0\leq y \leq 1$ is divided on
$M \times M$ equal cells. One trajectory is taken in the chaotic 
component (e.g. at $x_0=0.1/2\pi$, $y_0=0.1/2\pi$)
and is iterated on a large number of map iterations $t$, e.g.
$t=10^{12}$.  Then the UPFO matrix is defined as
$S_{ij} = n_{ij}/\sum_l n_{lj}$ where $n_{ij}$ is the number of 
transitions of the trajectory from a cell $j$ to a cell $i$.
By the construction we have $\sum_i S_{ij} =1$
and hence this UPFO  $S_{ij}$ belongs to 
the class of Perron-Frobenius operators. 
In this construction a  trajectory
visits only those cells which belong to one connected
chaotic component. Therefore the noise induced
by the discretization of the phase space
does not lead to a destruction of invariant curves,
in contrast to the original Ulam method \cite{ulam}
which uses all cells in the available phase space.
Since the trajectory is generated by a continuous map it cannot
penetrate inside the stability islands and on the physical 
level of rigor one can expect that, due to ergodicity
of dynamics on one connected chaotic component,
the UPFO constructed in such a way should converge
to the Perron-Frobenius operator of the continuous map
on a given subspace of chaotic component.

A mathematical prove of such a generalized Ulam conjecture
of the convergence of the UPFO built from one trajectory 
is not an easy task. Therefore, we performed extensive
numerical simulations which confirm the conjecture.
With this aim we checked that the results for the spectrum and
eigenstates of $S$ remain stable when $t$ is changed from 
$t=10^{10}$ to $10^{12}$, when we take another trajectory
in the same chaotic component, and when the size $M$
is increased (see detailed discussion below). 
To reduce the matrix size of $S_{ij}$ we use
the symmetry property of the map (\ref{eq_stmap})
which remains invariant under the transformation
$x \rightarrow 1-x$, $y \rightarrow 1 - y $
so that we can consider cells only in the 
lower half square with $0 \leq x \leq 1$, $0 \leq y \leq 1/2$
which contains $M^2/2$ cells.
At $K=K_g$ we find that the number of
cells visited by trajectory in this half square
scales as
$N_d \approx C_d M^2/2$ with $C_d \approx 0.42$.
This means that the chaotic component contains
about $40\%$ of the total area that is in a good agreement 
with the know result of \cite{chirikov1979}.

We used values of $M$ in the range $25\le M\le 1600$. To be more precise, 
for practical reasons, we determined the UPFO (actually the integer 
numbers $n_{ij}$) for the 
two largest values $M=1600$ and $M=1120$ by iterating a single trajectory 
as described above and for smaller values of $M$ we used an exact 
renormalization scheme by merging four neighbored cells (for a certain 
value of $M$) into one single cell (for $M/2$). In this way we obtained 
in an efficient way the UPFO also for smaller 
values $M=800,\,560,\ldots,\,35,\,25$ without the necessity to reiterate 
the same classical trajectory. 

For $t=10^{12}$ and $M=1600$ we have about 
$n_c \approx 2 t/(C_d M^2) \approx 1.8 \times 10^6$
transitions for each cell. This number is rather large
and  relative statistical fluctuations
are on a small level of $1/\sqrt{n_c} \sim 10^{-3}$. 

The direct exact diagonalization of the matrix $S$
can be done by standard computer routines 
which require memory resources of 
$N_d^2 \sim M^4$ double precision registers.
The computational time scales at $N_d^3 \sim M^6$.
Thus, for the map at $K=K_g$ we are  
practically limited to $M=280$ (with $N_d=16609$) 
as the maximum size for the 
full diagonalization. At such $M$ the statistical 
error is of the level $1/\sqrt{n_c} \sim 10^{-4}$.
Larger values of $M$ can be reached 
by the Arnoldi method as it is discussed in the next Section.

The eigenvalues $\lambda_j$ and corresponding
right eigenvectors $\psi_j(i)$ are defined from the equation
\begin{equation}
\label{eq_eigen}
\sum_{i=0}^{N_d-1} S_{mi} \psi_j(i) = \lambda_j\psi_j(m)  \; .  
\end{equation}
According to the Perron-Frobenium theorem \cite{mbrin}
we have the maximal eigenvalue $\lambda_0=1$
with the corresponding eigenstate $\psi_0(i)$
shown in Fig.~\ref{fig1} for four values of $M$.
All values $\psi_0(i)$ are non-negative in the agreement with the
theorem and have the meaning of the probabilities in a given cell
$i$. With the increase of $M$ the state $\psi_0(i)$
converges to a homogeneous ergodic measure on the chaotic component.
The stability islands are well incorporated inside
the chaotic component.

\begin{figure}[h!]
\begin{center}
\includegraphics[width=0.48\textwidth]{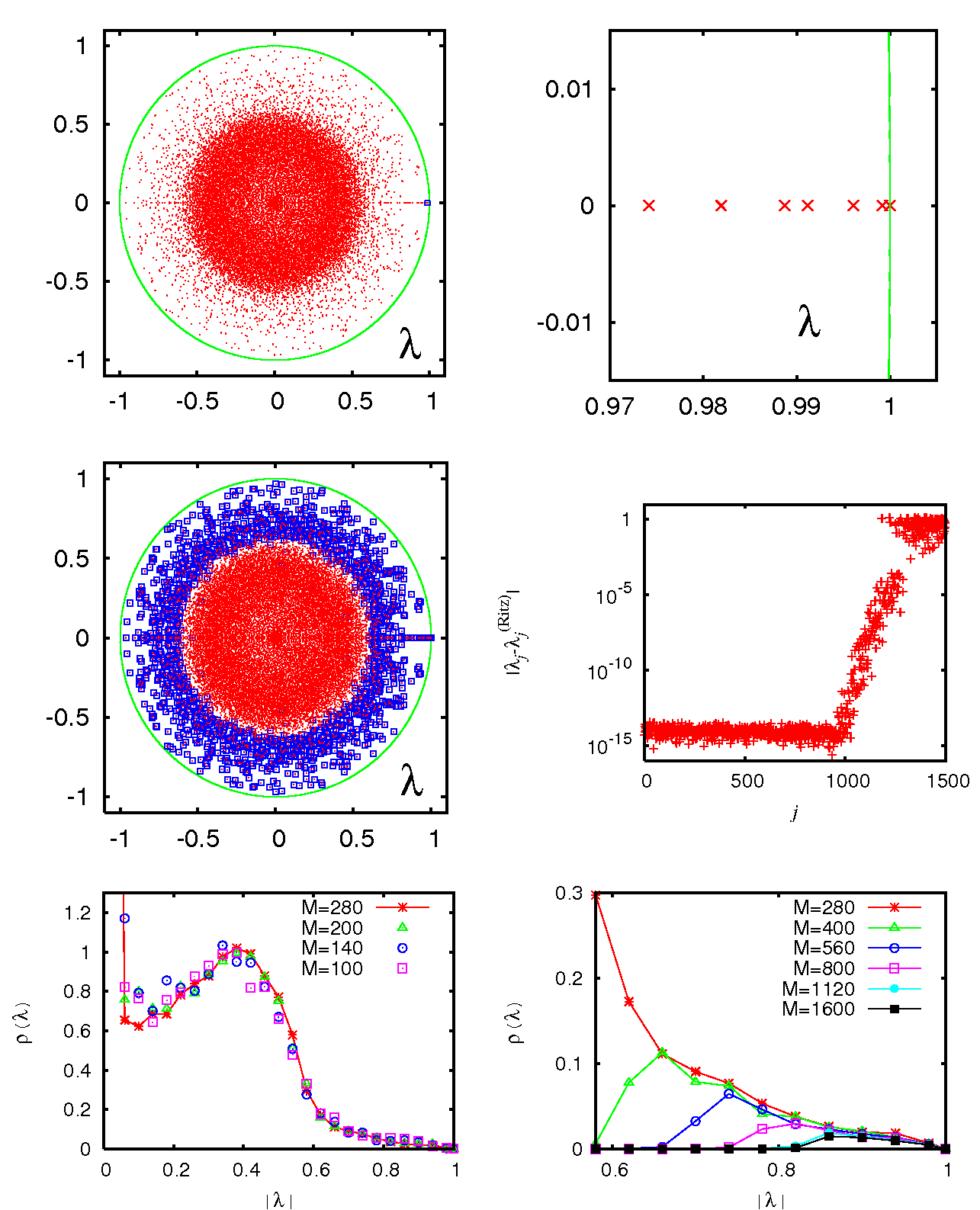}
\end{center}
\caption{\label{fig2} (Color online) Spectrum $\lambda_j$ 
of the UPFO of the map (\ref{eq_stmap}) at $K=K_g$.
{\em First row}~: The left panel shows the eigenvalue spectrum 
in the complex plane for $M=280$ and $N_d=16609$ 
by red/grey dots. The small blue/black square 
close to the region $\lambda=1$ 
is shown in more detail in the right panel 
with eigenvalues as red crosses. The green/grey curve 
represents the circle $|\lambda|=1$.
{\em Second row}~: 
In the left panel the Ritz eigenvalues (blue/black squares), 
obtained by the Arnoldi method for $M=280$ and with the Arnoldi 
dimension $n_A=1500$, are compared 
with the exact eigenvalues (red/grey dots). 
The right panel shows the modulus of the differences between the 
exact eigenvalues and the Ritz eigenvalues as 
a function of the level number
$j$ with eigenvalues sorted by decreasing modulus~: $|\lambda_0|=1>
|\lambda_1|>|\lambda_2|>\cdots$. The Ritz eigenvalues are 
numerically correct (with an error $\sim 10^{-14}$) for more then 
1000 first eigenvalues thus demonstrating 
the very good convergence of the Arnoldi method.
{\em Third row}~: The left panel shows the density $\rho(\lambda)$
of eigenvalues in the complex plane, being normalized 
by $\int \rho(\lambda)\,d^2\lambda = 1$, as a function of the modulus 
$|\lambda|$ for the values $M=100,\,140,\,200,\,280$. 
The peak  at $|\lambda|=0.02$ is outside the plot range 
and has values $\rho(0.02)=7.7$ $(M=280)$, 
$8.3$ $(M=200)$, $9.0$ $(M=140)$, and $10$ $(M=100)$. 
The right panel shows the density $\rho(\lambda)$
in the region $|\lambda|\in[0.58,\,1]$ for $M=280$ 
(full spectrum) and 
$M=400,\,560,\,800,\,1120,\,1600$ (partial spectrum). 
For $400\le M\le 1120$ only the largest 3000 eigenvalues 
and for $M=1600$ the largest 5000 eigenvalues were 
calculated by the Arnoldi method and therefore 
the corresponding densities deviate from the convergent  
density curve at small values of $\lambda$.
}
\end{figure}

Another confirmation of the convergence of the UFPO
in the limit of large $M$ is presented in Fig.~\ref{fig2}.
In a first approximation the spectrum $\lambda$ of $S$
is more or less homogeneously distributed
in the polar angle $\varphi$ defined as
$\lambda_j = |\lambda_j| \exp(i\varphi_j)$
(see left column of Fig.~\ref{fig2}). The two-dimensional 
density of states $\rho(\lambda)$ clearly 
converges to a limiting curve. This density of states is normalized by 
$\int \rho(\lambda)\,d^2\lambda = 1$ (for a full spectrum of $N_d$ 
eigenvalues). It 
drops when $|\lambda|$ approaches to $1$
but even at $|\lambda| \approx 0.9$
the convergence to a limiting curve is 
clearly seen. This is also confirmed by data with $400\le M\le 1600$ obtained 
from the Arnoldi method (which corresponds to a partial 
spectrum of $3000-5000\ll N_d$ eigenvalues with largest $|\lambda_j|$ and 
is therefore not properly normalized).

The convergence of $\rho(\lambda)$
at $N_d \rightarrow \infty$ implies
that the spectrum has a usual dimension
$d/2=1$ corresponding to the dimension of the phase space.
We note that the situation becomes different for dissipative maps
where the fractal Weyl law determines the number of
states in a given area of $\lambda$ that grows slower than $N_d$ 
(see \cite{zhirov,ermann1} and Refs. therein).
Our direct computation
of the number of states $N_\lambda$
in an interval $0.1 \leq \lambda \leq 1$ 
gives a linear dependence $N_\lambda \propto N_d$.

The properties of $\lambda_j$, with $|\lambda|$
being close to $1$ 
(see e.g. right top panel of Fig.~\ref{fig2}),
and their scaling with $M$
will be discussed in next Sections after a description
of the Arnoldi method which is especially efficient
in the computation of such eigenvalues.  

Let us note that our special checks
show that the variations of $\lambda_j$
with the change of initial trajectory
or its length $t$ remain on the level of statistical
accuracy $1/\sqrt{n_c}$. In the following 
we  present data obtained with the trajectory length
$t=10^{12}$ which is close to a maximal 
computational effort used in \cite{chirikov1999}
where $t \leq 10^{13}$ was used for the computation
of the Poincar\'e recurrences. Such a large value
of $t$ allows for the trajectory to penetrate into the very close
vicinity of the critical invariant curve
that becomes important at large $M$.

%%%%%%%%%%%%%%%%%%%%%%%%%%%%%%%%%%%%%%%%%%%%%%%%%%%%%%%%%
\section{Arnoldi method} 

In order to capture features of small 
phase space structures (such as small stable islands) and to get 
a better approximation of the continuous limit ($M\to\infty$) it is of 
course desirable to increase $M$ further than the value
$M=280$ accessible by the exact diagonalization. 
Fortunately the 
matrix $S$ is very sparse with an average number 
of non-zero connecting elements per row 
(or per column) being $\kappa_c \approx 5$  (and maximum 
number of links $\kappa_m=6$, at $K_g$) and 
$12$ (and $\kappa_m =20$ at $K=7$). The value of a maximal number
of non-zero elements is determined by a local stretching
given by the monodromy matrix, thus we have approximately
$\kappa_c \sim \exp(h)$ where $h$ is the Kolmogorov-Sinai entropy
(see \cite{chirikov1979} and discussion in \cite{zhirov}).
Since $\kappa_c \ll N_d$ we can calculate and store 
the matrix $S$ for larger values of $M$ and also effectively 
compute the product of $S$ with an arbitrary 
vector with $\kappa_c \times N_d$ operations. 

Furthermore, we 
are primary interested in the part of the spectrum with eigenvalues 
of modulus $|\lambda_j|$ close to $1$, or in other words with 
minimal decay rates $\gamma_j=-2\ln(|\lambda_j|)$, in order to capture 
the long time properties of chaotic dynamics with the UPFO iterations. 

We have therefore used the 
Arnoldi method \cite{book_with_arnoldi_method} which is perfectly 
adapted for this situation. This method is similar in spirit to the 
Lanzcos method, but is adapted for non-hermitian 
or non-symmetric matrices. It has 
allowed us to compute a considerable number of eigenvalues (with 
largest modulus) and the associated eigenvectors 
of $S$ for the values $M=400,\,560,\,800,\,1120,\,1600$ 
corresponding to the matrix dimension of the UPFO 
$N_d=33107,\,63566,\,$ $127282,\,245968,\,494964$ 
(for the map (\ref{eq_stmap}) at $K_g$) 
which are absolutely inaccessible 
by a full matrix diagonalisation. 
For the case with strong chaos at $K=7$ or the 
separatrix map the matrix dimension is even close to $N_d\approx 10^6$ 
for $M=1600$. In order to provide for a self-contained presentation, 
we give a short description of this method here.

The main idea of the Arnoldi method is 
to construct a subspace of ``modest'', 
but not too small, dimension $n_{\rm A}$ (in the following called the 
{\em Arnoldi-dimension}) generated by the vectors 
$\xi_0,\,S\xi_0,\,S^2\xi_0\,\ldots\,S^{n_{\rm A}-1}\xi_0$ 
(called {\em Krylov space}) where $\xi_0$ 
is some normalized initial vector and 
to diagonalize the projection of $S$ onto this subspace. The resulting 
eigenvalues are called the {\em Ritz eigenvalues} which represent often 
very accurate approximations of the exact eigenvalues of $S$, at least 
for a considerable fraction of the Ritz eigenvalues with largest modulus.

To do this more explicitly, we first construct 
recursively an orthonormal set 
(of $n_{\rm A}+1$ vectors) $\xi_0,\xi_1,\ldots\,,\xi_{n_{\rm A}}$. 
For $k=0,\,1,\ldots,\,n_{\rm A}-1$ we define the vector 
$v_{k+1}$ as the Gram-Schmidt orthogonalized (but not yet normalized) 
vector of $S\,\xi_k$ 
with respect to $\xi_0,\ldots\,,\xi_k$ and store the matrix 
elements $h_{j,k}=\<\xi_j\,|S|\,\xi_k\>$ for $j=0,\ldots\,,k$
which were used during the orthogonalization scheme. 
Furthermore we define the matrix element 
$h_{k+1,k}=\parallel v_{k+1}\parallel$ 
and normalize $v_{k+1}$ by $\xi_{k+1}=v_{k+1}/h_{k+1,k}$. Then the 
product $S\,\xi_k$ can be expressed in terms 
of the orthonormal vectors $\xi_j$ by:
\begin{equation}
\label{eq_Sxi_expand}
S\,\xi_k=\sum_{j=0}^{k+1} h_{j,k}\,\xi_j
\end{equation}
and therefore the matrix $h_{j,k}$ is the representation matrix 
of $S$ in the Krylov space. This expansion is called in the 
mathematical literature \cite{book_with_arnoldi_method} 
{\em Arnoldi-decomposition} when written in matrix form and it is 
actually an exact identity. However, it 
is not closed since $S\,\xi_k$ requires a contribution 
of $\xi_{k+1}$ unless $h_{k+1,k}=0$ for some value of $k$ in which 
case we would have obtained an exact $S$-invariant 
subspace and the diagonalization 
of the {\em Arnoldi matrix} $h_{j,k}$ would provide 
a subset of  exact eigenvalues of $S$ 
(those with eigenvectors in the $S$-invariant subspace).
 An interesting 
situation appears if due to numerical rounding errors $h_{k+1,k}$ is 
very small and not exactly zero. Then the method automatically 
generates, with the help of rounding errors, 
a new ``pseudo-random'' start 
vector and explores a new subspace orthogonal 
to the first $S$-invariant 
subspace which is actually useful to obtain further eigenvalues. 

However, when diagonalizing the UPFO $S$ for a chaotic map with large 
dimension this situation, which 
may be quite important in certain other cases, 
does not happen and $h_{k+1,k}$ is always different 
from zero (actually $h_{k+1,k}$ is quite comparable 
in size to the modulus of 
eigenvalue $\lambda_k$). Therefore we have to cut 
the above iteration at 
some maximal value of $k$. In order to calculate the Arnoldi matrix of 
dimension $n_{\rm A}$ one must actually be careful to determine 
$n_{\rm A}+1$ vectors, otherwise 
one would miss the last column of the matrix $h$. We also note that 
the Arnoldi matrix $h_{j,k}$ is of Hessenberg form 
($h_{j,k}=0$ if $j>k+1$) 
which simplifies the numerical diagonalization 
since one can directly call the 
subroutine for the $QR$-diagonalization 
and omit the first, quite expensive, 
step which transforms a full matrix to Hessenberg form by Householder 
transformations. 

We mention as a side remark that for symmetric or 
hermitian matrices $S$ 
one can show that the matrix $h_{j,k}$ is 
tridiagonal and the orthogonalization needs only 
to be done with respect to 
the last two vectors resulting in the well known Lanczos algorithm. 
In principal, the use of an exact mathematical property, 
which may be violated 
due to numerical rounding errors, is somewhat tricky 
and may require special 
treatment in the various variants of the Lanczos method. 
However, the Arnoldi method always requires 
orthogonalization with respect to {\em all} previous vectors and does 
not suffer from this kind of problem but it 
is also more expensive than the Lanczos algorithm. 

The Arnoldi method requires $N_d\, \kappa_c$ 
double precision registers 
to store the non-zero matrix elements of $S$, 
$N_d\,n_{\rm A}$ registers to store the vectors 
$\xi_k$ and const.$\times n_{\rm A}^2$ registers to store 
$h_{j,k}$ (and various copies of $h$). The computational time 
scales as $N_d\,\kappa_c \,n_{\rm A}$ 
for the computation of $S\,\xi_k$, with 
$N_d\,n_{\rm A}^2$ for the Gram-Schmidt 
orthogonalization procedure (which is typically dominant) and with 
const.$\times n_{\rm A}^3$ for the diagonalization of $h_{j,k}$. 

In the practical applications of the Arnoldi method an important point 
concerns the ``good'' choice of the initial vector $\xi_0$. It 
is actually a bad idea to chose a vector which is close 
to the eigenvector of 
maximal eigenvalue (or other eigenvalues) because this would suppress 
contributions of other eigenvectors which we want to retain. 
A much better 
choice is a random initial (normalized) vector. 
During the Arnoldi iteration 
the method will automatically suppress 
the eigenvector contributions with 
respect to the smallest values $|\lambda_j|$ and 
retain the contributions 
of eigenvalues close to the unit circle. 
If the spectrum has some well-defined 
modest gap between $\lambda_0=1$ and the other eigenvalues the random 
initial vector is indeed a very good choice and we have used this choice 
for the case of map (\ref{eq_stmap})  at $K=7$ which we 
discuss in Section 5. 
However, at critical $K_g$ there is no real gap 
(see for example the upper right panel in Fig. \ref{fig2}) 
and there is also a considerable number 
of eigenvalues close to unit circle. 
In this case the inherent suppression 
of small eigenvalues by the Arnoldi 
method may not be sufficiently fast, if $|\lambda_k|^k$ is not small for 
$k$ close to the chosen Arnoldi dimension $n_{\rm A}$. Therefore we have 
chosen here an initial (normalized) vector obtained from an initial 
number of iterations of $S$ applied to a random vector~: 
$\xi_0\propto S^{n_{\rm ini.}}\,\xi_{\rm random}$ with $n_{\rm ini.}$ 
being the number of initial $S$-iterations which we have chosen 
to scale with $M^2$: $n_{\rm ini.}=M^2/200$ 
(except for the case $M=1600$ where we have 
chosen $n_{\rm ini.}=7000$).

As a first illustration, we have applied the Arnoldi method with 
$n_{\rm A}=1500$ to the case of $M=280$ and $N_d=16609$ for the Chirikov 
standard map at $K_g$, for which we were still able to diagonalize the 
full matrix $S$. In the middle right panel of 
Fig. \ref{fig2}, we show the modulus of the difference 
(in the complex plane) 
of the Ritz eigenvalues and the exact eigenvalues as a function 
of the level number $j$. 
The first $1000$ Ritz eigenvalues, out of $1500$ 
in total, are numerically correct with a deviation $\sim 10^{-14}$ 
entirely due to 
numerical rounding errors. If we only require graphical precision 
($\sim 10^{-5}$) there are actually $1200$ Ritz eigenvalues which 
are still acceptable. This can also be seen in the middle left panel of 
Fig. \ref{fig2} where we compare the spectrum in 
the complex plane of the full 
matrix $S$ with the partial spectrum obtained by the Arnoldi method. 
This provides a quite impressive confirmation of 
the accuracy of the Arnoldi 
method. For larger values of $M$ we have done similar verifications, for 
example by comparing the Ritz eigenvalues 
for different values of $n_{\rm A}$ 
or for different initial vectors. Choosing typically $n_{\rm A}=1500$ or 
$n_{\rm A}=3000-5000$ for the largest values of 
$M=1120,\,M=1600$, we always have a considerable number (at least 
$500$ to $1000$) of numerically accurate eigenvalues. 

Concerning the (right) eigenvectors, we prefer to determine 
them independently by the method of inverse 
vector iteration which provides numerical reliable (real or complex) 
eigenvectors with 
$n_{\rm A}^2$ operations per eigenvector 
due to the Hessenberg form of $h_{j,k}$. 
Suppose that $\varphi$ is such an eigenvector 
of $h_{j,k}$ with eigenvalue $\lambda$,
\begin{equation}
\label{eq_eigvect_h}
\lambda\varphi_j=\sum_{k=0}^{n_{\rm A}-1}\,h_{j,k}\,\varphi_k\ ,
\end{equation}
then we obtain by Eq. (\ref{eq_Sxi_expand}), the corresponding 
eigenvector $\psi$ of $S$ directly from:
\begin{equation}
\label{eq_eigvect_S}
\psi=\sum_{k=0}^{n_{\rm A}-1}\,\varphi_k\,\xi_k\ .
\end{equation}

%%%%%%%%%%%%%%%%%%%%%%%%%%%%%%%%%%%%%%%%%%%%%%%%%%%%%%%%%
\section{Chirikov standard map at $K_g$}

We first apply the Arnoldi method to the Chirikov standard map 
(\ref{eq_stmap}) at the critical value $K_g=0.971635406$. 
The Arnoldi method allowed 
us to obtain a considerable number of eigenvalues and eigenvectors of the 
UPFO $S$ for the values $M=400$, $560$, $800$, $1120$, $1600$ 
corresponding to the matrix dimension 
$N_d=33107$, $63566$, $127282$, $245968$, $494964$. We choose 
the Arnoldi dimension $n_{\rm A}=3000$ for $M\le 1120$ and 
$n_{\rm A}=5000$ for $M\le 1600$ and we also compute the first $500$ 
eigenvectors for each case.
Even though we are not able to calculate 
the full spectrum for these cases, the partial densities 
of the eigenvalues 
in the complex plane, for $|\lambda|$ close to $1$, 
are in a good agreement
with the full densities obtained for $M\le 280$ 
as can be seen in the bottom right panel of 
Fig. \ref{fig2}. Below we present
the most important part of obtained data, more details
with many eigenstates and high resolution figures are 
available at \cite{qwlib}.

\begin{figure}[h]
\begin{center}
\includegraphics[width=0.48\textwidth]{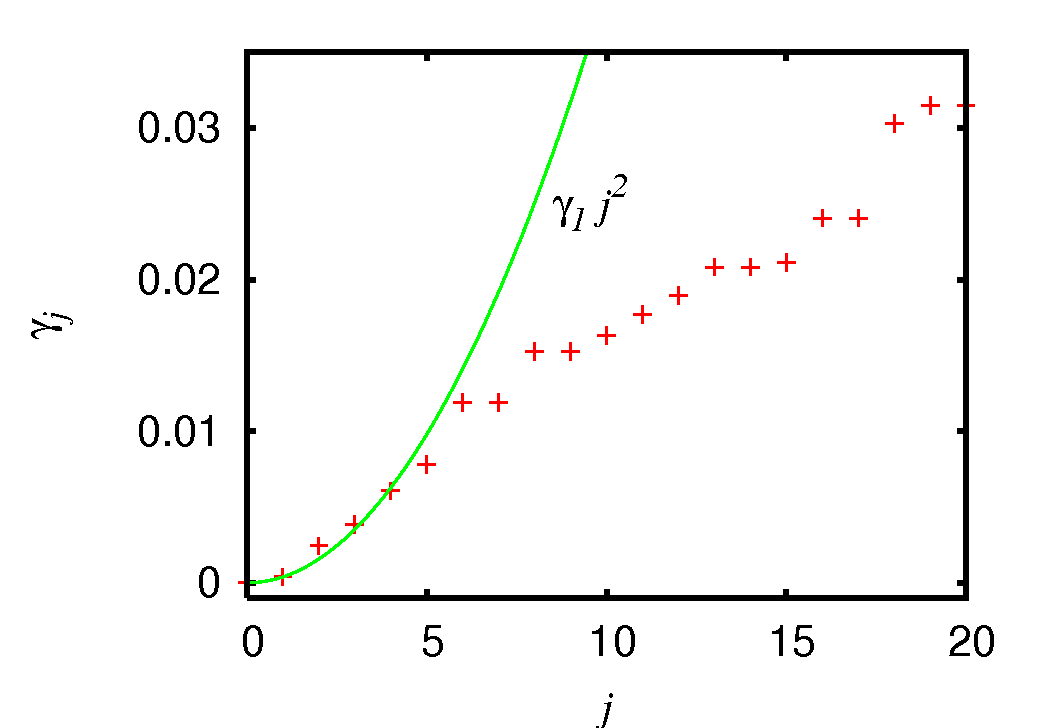}
\end{center}
\caption{\label{fig3} (Color online) 
Decay rates $\gamma_j=-2\ln(|\lambda_j|)$ versus level number $j$ (red 
crosses) for the UPFO eigenvalues $\lambda_j$ of the map (\ref{eq_stmap})
at $K=K_g$,  $M=800$ and $N_d=127282$. 
The green curve corresponds to the 
quadratic dispersion law 
$\gamma_j\approx \gamma_1\,j^2$ which is approximately valid for the 
diffuson modes with $0\le j\le 5$. 
}
\end{figure}

In Fig. \ref{fig3} we show the decay rates $\gamma_j=-2\ln(|\lambda_j|)$ 
as a function of the level number $j$ 
(with eigenvalues sorted by decreasing 
$|\lambda_j|$ or increasing $\gamma_j$) for the case $M=800$. We note 
that the first 6 eigenvalues follow quite closely 
a quadratic dispersion law 
$\gamma_j\approx \gamma_1\,j^2$ for $0\le j\le 5$. These $6$ eigenvalues 
are actually real, positive and close to 1. 
Their corresponding eigenvectors 
(which are also real) extend over the full phase space covered 
by the chaotic trajectory used to determine the UPFO. 

\begin{figure}[h]
\begin{center}
\includegraphics[width=0.48\textwidth]{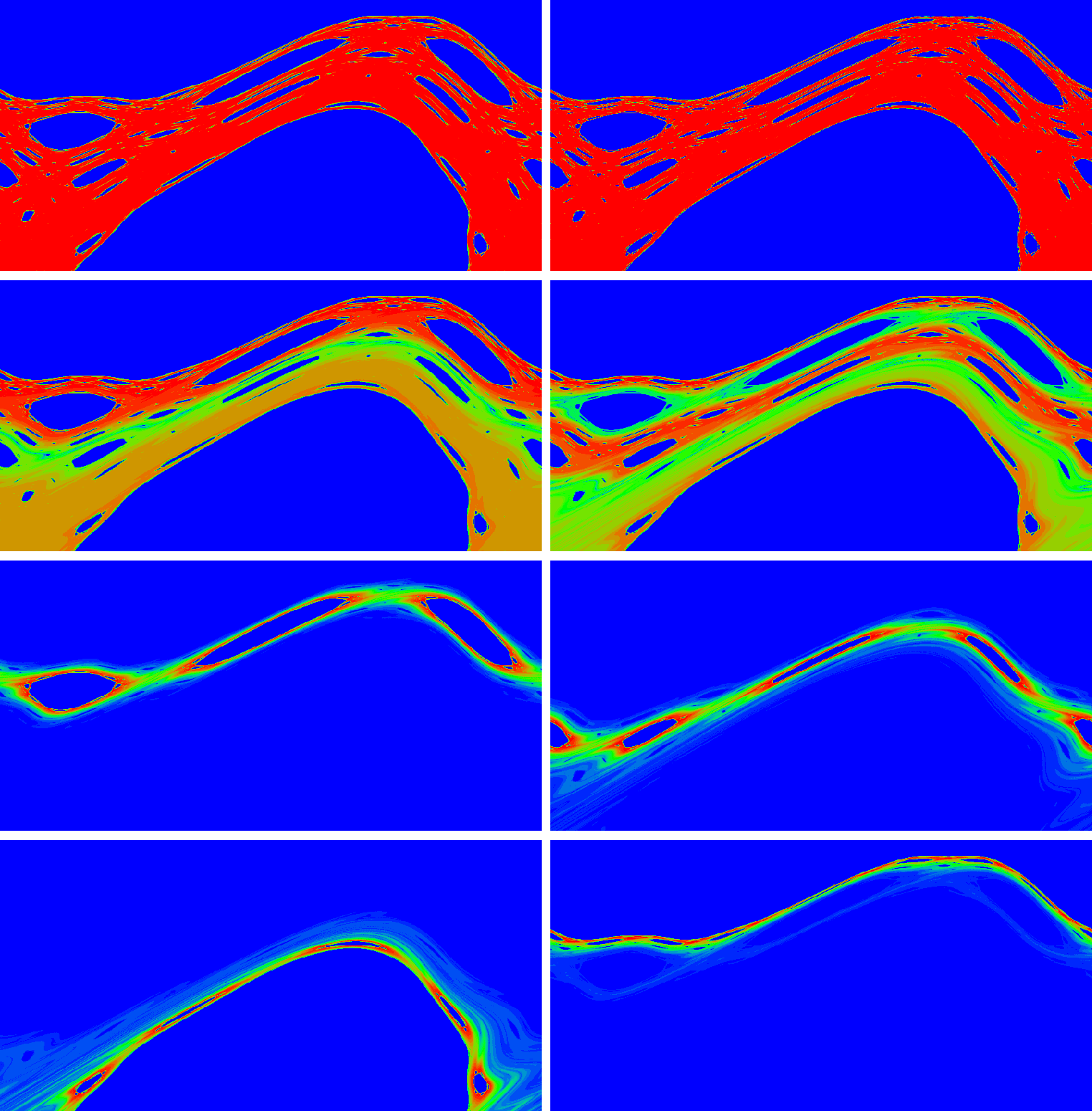}
\end{center}
\caption{\label{fig4} (Color online)
{\em First row}~: Density plot of the eigenvector with eigenvalue 
$\lambda_0=1$ for $M=800$ and $N_d=127282$ (left panel) and 
for $M=1600$ and $N_d=494964$ (right panel). In last three rows 
$M=800$ and $N_d=127282$. 
{\em Second row}~: Density plot of the modulus of the components of the 
eigenvectors for $\lambda_1=0.99980431$ (left panel) and 
$\lambda_2=0.99878108$ (right panel). 
{\em Third row}~: Density plot of the modulus of the components of the 
eigenvectors for 
$\lambda_6=-0.49699831+i\,0.86089756\approx |\lambda_6|\,e^{i\,2\pi/3}$ 
(left panel) and 
$\lambda_8=0.00024596+i\,0.99239222\approx |\lambda_8|\,e^{i\,2\pi/4}$
(right panel). 
{\em Fourth row}~: Density plot of the modulus of the components of the 
eigenvectors for 
$\lambda_{13}=0.30580631+i\,0.94120900\approx |\lambda_{13}|\,e^{i\,2\pi/5}$ 
(left panel) and 
$\lambda_{19}=-0.71213331+i\,0.67961609\approx |\lambda_{19}|
\,e^{i\,2\pi(3/8)}$
(right panel). 
}
\end{figure}

In  the first row of Fig. \ref{fig4} 
we show the density plots of the (right) eigenvector 
$\psi_0(m)$ in phase space representation (the index $m$ 
gives the discretized phase space 
position of the cell $m$) for $M=800$ and $M=1600$. 
The second row shows $|\psi_1(m)|$ and $|\psi_2(m)|$ for $M=800$. 
In agreement with the ergodic theorem, 
the  eigenvector $\psi_0$ represents a nearly 
uniform density on the chaotic component. 
The eigenvectors $\psi_j$ 
for $1\le j\le 5$ 
(and also for certain higher values 
$j \leq 15$ if the associated eigenvalue is 
real, positive and close to $1$) 
correspond to some kind of ``diffuson modes'' 
with a roughly uniform distribution in the angle coordinate $x$
and a wave structure 
with a finite number of nodes in the action coordinate $y$. 
For example $|\psi_1(m)|$ 
is maximal at the upper and lower borders of the available phase space  and 
$\psi_1(m)$ changes sign on exactly one curve in between. 
For $|\psi_2(m)|$ 
there are three maximal curves and two node curves with a sign change of 
$\psi_2(m)$ and so on for other diffuson modes. 
Such diffuson modes with quadratic spectrum
naturally appears as a solution of the diffusion equation 
\begin{equation}
\label{eq_dif}
\frac{\partial \rho}{\partial t} = 
\frac{\partial }{\partial y}
\left( D_y\frac{\partial \rho}{\partial y} \right) \; ,
\end{equation}
with boundary conditions $\partial \rho/\partial y =0$
at $y=0$ and $y \approx 1-r_g$: 
$\rho_j(y) \propto \cos(\pi j y/(1-r_g))$, 
$\gamma_j \approx \pi^2 D_y j^2/(1-r_g)^2$ (assuming $D_y$ to be constant 
on the interval $0\le y \le 1-r_g$).

The structure of the eigenvectors for complex eigenvalues 
(or real negative 
eigenvalues close to ``$-1$'') is very different and corresponds to 
the ``resonance modes'' which are typically concentrated 
(or even localized) around one (or a chain of few) 
resonance(s). This can be seen in the third and fourth 
rows of Fig.~\ref{fig4} containing 
the density plots of $|\psi_j(m)|$ for $j=6$ and $j=8$ (third row) 
and for $j=13$ and $j=19$ (fourth row). The complex phase 
$\varphi_j$ of $\lambda_j=|\lambda_j|\,e^{i\,\varphi_j}$ 
for such an eigenvalue represents quite well the periodicity of a 
trajectory with a period 
$q$ if $\varphi_j\approx 2\pi(p/q)$ is approximated by a rational number 
times $2\pi$.  The fraction $p/q$ 
represents the position of the resonance in the rotation number $r$.
In Fig. \ref{fig4} we can identify $\varphi_6\approx 2\pi(1/3)$, 
$\varphi_8\approx 2\pi(1/4)$, $\varphi_{19}\approx 2\pi(3/8)$ 
and as a secondary resonance (close to the main resonance at $p=0$) 
$\varphi_{13}\approx 2\pi(2/5)$. 

The density plots of further resonance modes 
(with complex or real negative 
eigenvalue) are typically similar and they 
approximately repeat  the modes associated with 
the largest resonances but with modified 
phases and decay rates. For example the  density of the 
mode $\lambda_{10}=-0.99187524$ (for $M=800$) with the phase 
$\varphi_{10}=2\pi(2/4)$ is very similar to the density of 
the mode $\lambda_8$ with $\varphi_8=2\pi(1/4)$ 
representing the resonance 
at $1/4$. Another example concerns the resonance modes 
close to the resonance $1/3$ 
(see density plot of the mode $\lambda_6$ in Fig. \ref{fig4}). 
There is a certain number of higher modes 
with phases that can be written 
in the form $2\pi(p_1/3+p_2/8)$ 
with certain small integer numbers $p_1, p_2$. 

\begin{figure}[h]
\begin{center}
\includegraphics[width=0.48\textwidth]{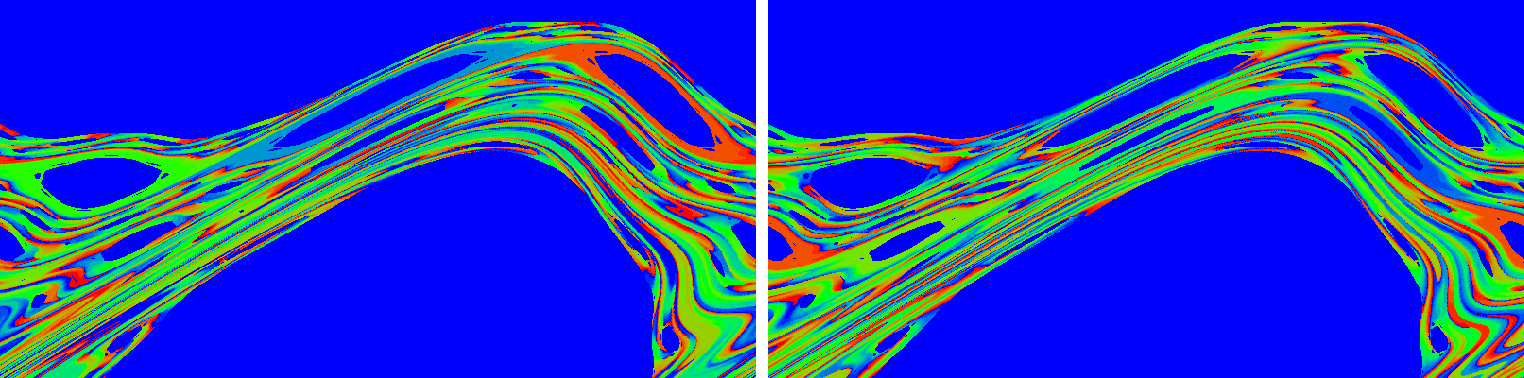}
\end{center}
\caption{\label{fig5} (Color online)
Density plot of the complex phase of the components of the 
eigenvectors for 
$\lambda_6=-0.49699831+i\,0.86089756\approx |\lambda_6|\,e^{i\,2\pi/3}$ 
(left panel) and 
$\lambda_8=0.00024596+i\,0.99239222\approx |\lambda_8|\,e^{i\,2\pi/4}$
(right panel). 
Deep blue corresponds to either empty cells or phase $=-\pi$, green to 
phase $=0$ and red to phase $=\pi$ ($M=800$ and $N_d=127282$).
}
\end{figure}

One may also ask the question in how far the complex phases of the 
eigenvector components carry interesting information. 
As a illustration we show in Fig. \ref{fig5} for two examples the density 
plot of the complex phase of 
$\psi_j(m)$ for $j=6$ (or $j=8$) and $M=800$. Even though these modes are 
quite well localized (see Fig. \ref{fig4}) close to the resonances 
$1/3$ (or $1/4$) 
they still extend, with well defined complex phases of $\psi_j(m)$, to 
the full accessible phase space  described by the UPFO. In the 
region with very small values of $|\psi_j(m)|$ 
the phase dependence is quite 
complicated and one cannot provide a simple physical interpretation. 
However, in the region of maximal $|\psi_j(m)|$ close to the classical 
resonance $1/3$ ($1/4$) for $j=6$ ($j=8$) one can identify a simple 
structure where the phase is roughly constant on a boundary layer outside 
of each stable region associated to the resonance but with different 
values $2\pi(l/3)+$const. ($2\pi(l/4)$+const.) for each of the three 
(four) islands characterized by the number $l=0,1,2$ ($l=0,1,2,3$).

\begin{figure}[h]
\begin{center}
\includegraphics[width=0.48\textwidth]{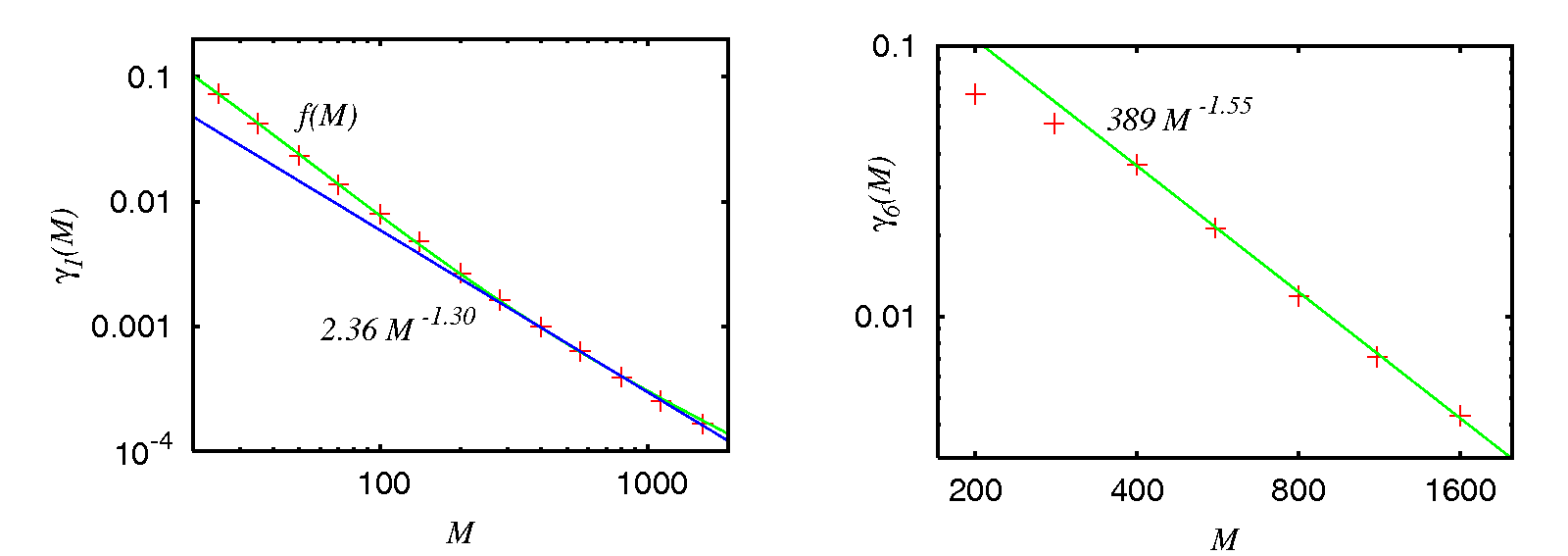}
\end{center}
\caption{\label{fig6} (Color online)
The left panel shows the decay rate 
$\gamma_1(M)$ of the first ``excited'' 
diffuson mode (see second row, left panel in Fig. \ref{fig4}) 
of the UPFO as a function of $M$ (red crosses) 
in a double logarithmic scale. The lower 
(blue) line corresponds to the power law fit 
$2.36\,M^{-1.30}$. The 
upper (green) curve corresponds to the fit~: 
$f(M)=\frac{D}{M}\,\frac{1+C/M}{1+B/M}$ 
with $D=0.245$, $C=258$ and $B=13.1$. 
The right panel shows the decay rate $\gamma_6(M)$ of the mode 
associated to the resonance $1/3$ (see third row, left panel 
in Fig. \ref{fig4}) as a a function of $M$ (red crosses) in a double 
logarithmic scale. The 
(green) line corresponds to the power law fit $389\,M^{-1.55}$. 
Both power fits were obtained for the range $400\le M\le 1600$ 
while for the fit with $f(M)$ all values $25\le M\le 1600$ were used.
}
\end{figure}

Concerning the first non-zero decay rates one important question is the 
dependencies of $\gamma_j(M)$ as a function of $M$ and in particular what 
happens in the limit $M\to\infty$. 
In Fig. \ref{fig6}, we show $\gamma_1(M)$, 
for the first non-trivial diffuson mode (left panel), and 
$\gamma_6(M)$, for the first resonance mode (right panel), versus $M$ 
in a double logarithmic scale. 
In both cases $\gamma_j(M)$ seems to 
tend to zero for $M\to\infty$ and a power law fit for 
the range $400\le M\le 1600$ indicates the behavior 
$\gamma_1(M)\approx 2.36\,M^{-1.30}$ and $\gamma_6(M)\approx 389\,M^{-1.55}$. 
However, the situation for $\gamma_1(M)$ 
seems more subtle and the curvature, 
when taking into account the range of 
all values $25\le M\le 1600$, seems to 
indicate a transition 
from $M^{-2}$ for small $M$-values to $M^{-1}$ for larger values
of $M$. Actually 
the data can also be well described by the fit: 
\begin{equation}
\label{eq_gamma1_fit}
\gamma_1(M)\approx f(M)=\frac{D}{M}\,\frac{1+C/M}{1+B/M}
\end{equation}
with $D=0.245$, $C=258$ and $B=13.1$. 

A physical interpretation of 
this behavior will be discussed  in Section 7. 
Here we only 
note that the vanishing limit $\lim_{M\to\infty}\gamma_j(M)=0$ is coherent 
with the observation that the relaxation to the uniform 
ergodic eigenvector 
is described by a power law decay of the Poncar\'e recurrences
in time. 

\begin{figure}[h]
\begin{center}
\includegraphics[width=0.48\textwidth]{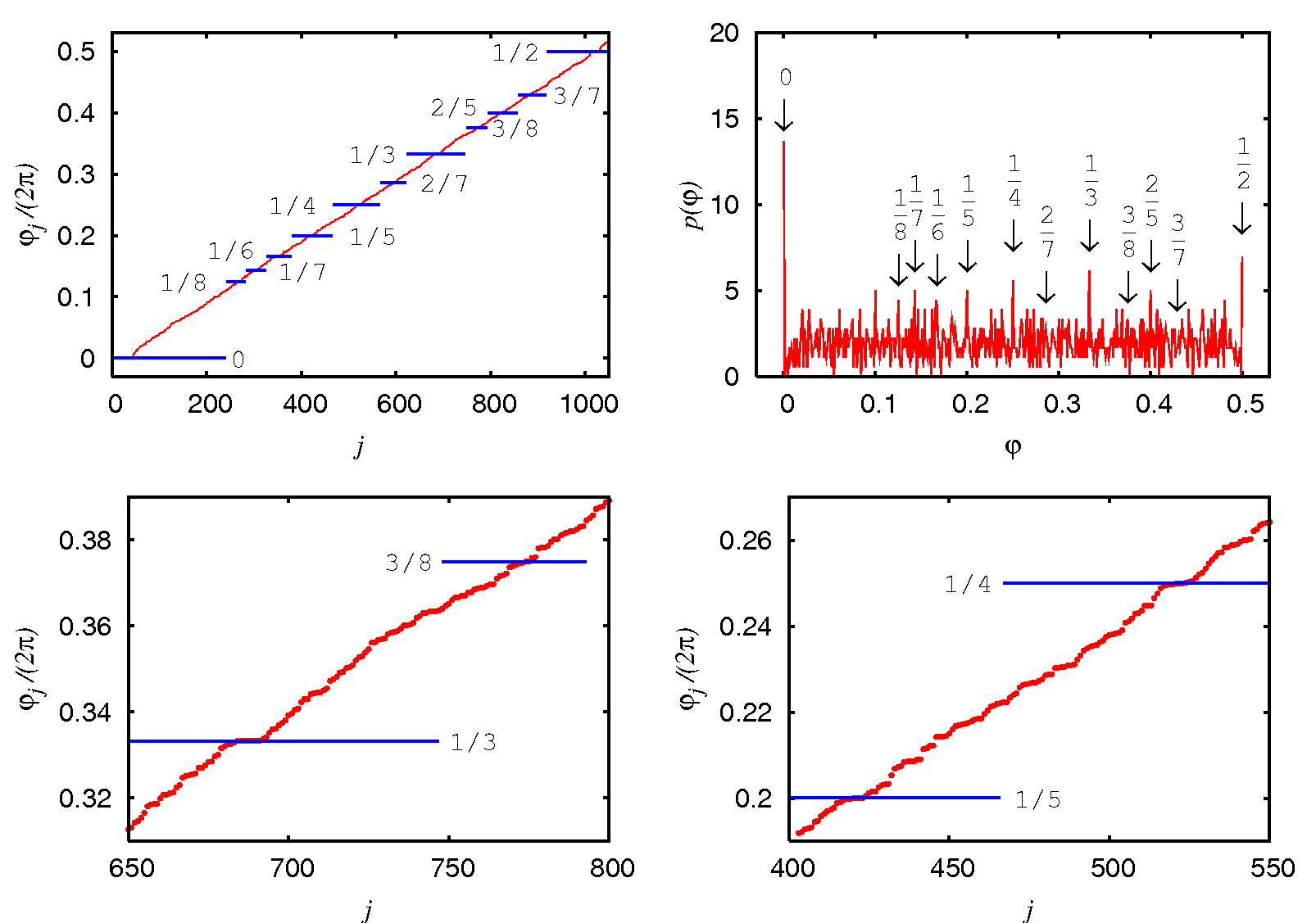}
\end{center}
\caption{\label{fig7} (Color online)
The top left panel shows the ordered complex phases $\varphi_j$ of the 
complex eigenvalues $\lambda_j=|\lambda_j|\exp(i\varphi_j)$
of the UPFO for $M=1600$ and the largest $2000$ eigenvalues 
(obtained by the Arnoldi method with the Arnoldi dimension $n_A=3000$) 
as a function of the index $j$ such that $\varphi_j \le \varphi_{j+1}$. 
The horizontal (blue) lines represent the fractional values 
of the Farey sequence of order $8$ corresponding to steps in the 
phase number function. The two bottom panels show the same data at 
with higher resolution  in a vicinity of  
ratios $1/3$ (left panel) or $1/4$ (right panel) 
for the phase rotation number.
The top right panel shows the distribution of 
the phase rotation numbers 
obtained by a histogram 
of bin width $1/840$ and the complex phases of the 
largest 3000 eigenvalues. 
The positions of the local maxima correspond quite well 
to the fractional values of the Farey sequence. 
Since the complex eigenvalues appear in 
complex conjugate pairs phase numbers $\varphi/2\pi>1/2$ 
have been mapped to values 
below $1/2$ by $\varphi/2\pi \to 1-\varphi/2\pi$. 
}
\end{figure}

Before we close this section, we come back to the discussion of the 
complex phases of the eigenvalues. As already observed above, 
it seems that 
for the resonance modes the eigenvalue phases $\varphi_j$ 
of $\lambda_j=|\lambda_j|\exp(i\varphi_j)$
are close to $2\pi\times p/q$, where $p/q$ is some 
rational number  with small values of $q$. 
In Fig. \ref{fig7} we 
analyze the statistical behavior of $\varphi_j/(2\pi)$ in two ways. 
Namely,
we show $\varphi_j$ versus an index $j$ ordered in such a way that 
$\varphi_j<\varphi_{j+1}$. 
The horizontal lines show the 
rational numbers of the Farey sequence 
of order $8$ (i.~e. all irreducible 
rational numbers between 0 and 1 with a maximal denominator $8$) 
obtained from a continuous fraction approximation of $\varphi_j/(2\pi)$. 
One can see that these rational values correspond to small steps 
indicating a larger than average probability to find a rational number 
with small denominator. This feature is seen even   clearer in the
top right panel of Fig. \ref{fig7} where
the distribution of $\varphi_j/(2\pi)$ has well defined peaks 
at the positions associated with the Farey sequence.

%%%%%%%%%%%%%%%%%%%%%%%%%%%%%%%%%%%%%%%%%%%%%%%%%%%%%%%%%
\section{Chirikov standard map at $K=7$}

We now turn to a particular case of strong chaos at $K=7$ which was previously 
studied in \cite{chsh} and by Chirikov in \cite{chirikov2000b}. 
According to \cite{chsh} the statistics of Poincar\'e recurrences
on line $y=0$ drops exponentially with time.
This is also a case even if one takes another line for recurrences,
e.g. $y=1/2$. In the latter case
the recurrences are mainly determined by a trajectory sticking
in a vicinity of two small stability islands
located  on a line $y=0$.
However, in this case the border
of islands is very sharp and 
the statistics of Poincar\'e recurrence still decays exponentially
up to rather long times $P(t)\sim \exp(-t/\tau)$ with a decay time $\tau=23.1$
\cite{chirikov2000b}. 

We have determined the UPFO $S$ at $K=7$ for the same values as previously 
($25\le M\le 1600$) using a 
trajectory of length $10^{11}$ which is sufficient due to the faster 
relaxation to the ergodic distribution (as compared to the case of 
critical $K_g$ where we used a trajectory of length $10^{12}$). Now 
the matrix size $N_d$ of $S$ is very close to its maximal value $M^2/2$ 
since with the exception of the two small stable islands nearly all cells are 
visited by the trajectory. For $M=1600$ we have $N_d=1279875$ .
\begin{figure}[h]
\begin{center}
\includegraphics[width=0.48\textwidth]{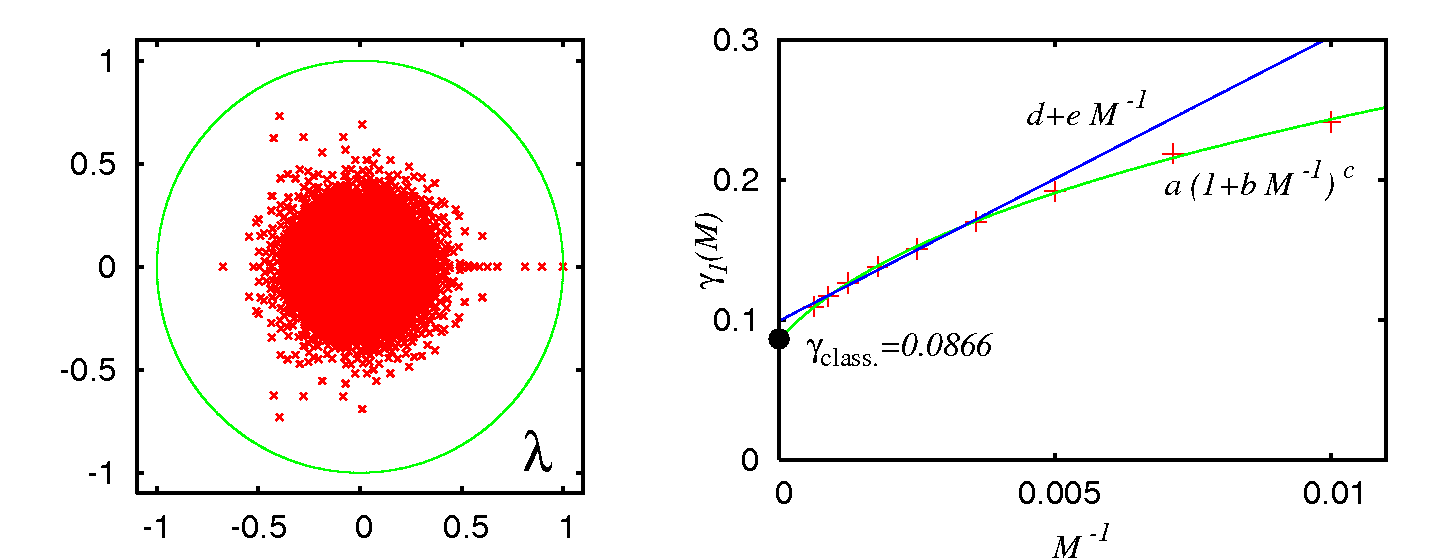}
\end{center}
\caption{\label{fig8} (Color online)
The left panel shows the eigenvalue spectrum in the 
complex plane of the UPFO ($t=10^{11}$ iterations) 
for the map (\ref{eq_stmap}) at $K=7$ 
for $M=140$ and $N_d=9800=140^2/2$ as red dots. The green curve 
shows the unit-circle $|\lambda|=1$.
The right panel shows the decay rate $\gamma_1(M)$ 
as a function of $M^{-1}$ for $100 \leq M \leq 1600$. The 
upper (blue) straight line corresponds to the fit 
$\gamma_1(M)=d+e\,M^{-1}$ for $M^{-1}\le 0.004$ resulting in 
$d=0.0994$ and $e=20.3$ suggesting the extrapolation 
limit $\lim_{M\to\infty} \gamma_1(M)=0.0994$. The lower (green) curve
corresponds to the fit $\gamma_1(M)=a\,(1+b\,M^{-1})^c$ 
for $M^{-1}\le 0.01$ resulting in $a=0.0857$, $b= 1370$ and 
$c=0.389$ suggesting the extrapolation 
limit $\lim_{M\to\infty} \gamma_1(M)=0.0857$. The black dot 
marks the decay rate $\gamma_{\rm cl}=0.0866$ found 
directly from the the Poincar\'e recurrences
in  \cite{chirikov2000b}. 
}
\end{figure}

We have calculated the full eigenvalue spectrum of $S$ by direct 
diagonalization for $M\le 140$ (corresponding to $N_d\le 9800$) and 
the first $n_{\rm A}=1500$ eigenvalues (and the first 500 eigenvectors) by the 
Arnoldi method for $M\ge 200$ ($N_d\ge 20000$). 
As compared to the case of critical $K_g$ the necessary time and memory 
resources are increased due to larger values of $N_d$ at given $M$ but 
on the other hand we get reliable eigenvalues for smaller numbers of 
the Arnoldi dimension because the modulus of the eigenvalues decay much faster 
with increasing level number. This can be seen at the eigenvalue spectrum 
in the complex plane shown in the left panel of Fig. \ref{fig8} for 
$M=140$. There are only few eigenvalues outside the circle of radius $0.5$ 
and we can also identify a clear gap between the first two eigenvalues 
$\lambda_0=1$ and $\lambda_1=0.8963823322$ (for $M=140$). 
The density of eigenvalues 
in the complex plane (normalized by $\int \rho(\lambda)\,d^2\lambda = 1$)
can be quite well approximated by the expression: 
\begin{equation}
\label{eq_densk7}
\rho(\lambda)\approx \exp\left(2.55-6.0\,|\lambda|-11.4\,|\lambda|^2\right)
\quad,\quad |\lambda|\le 0.65
\end{equation}
and for $|\lambda|>0.65$ we have $\rho(\lambda)< 0.001$. This expression 
fits the density for all values $25\le M\le 140$ for which we have been 
able to compute the full eigenvalue spectrum of $S$. 
\begin{figure}[h]
\begin{center}
\includegraphics[width=0.48\textwidth]{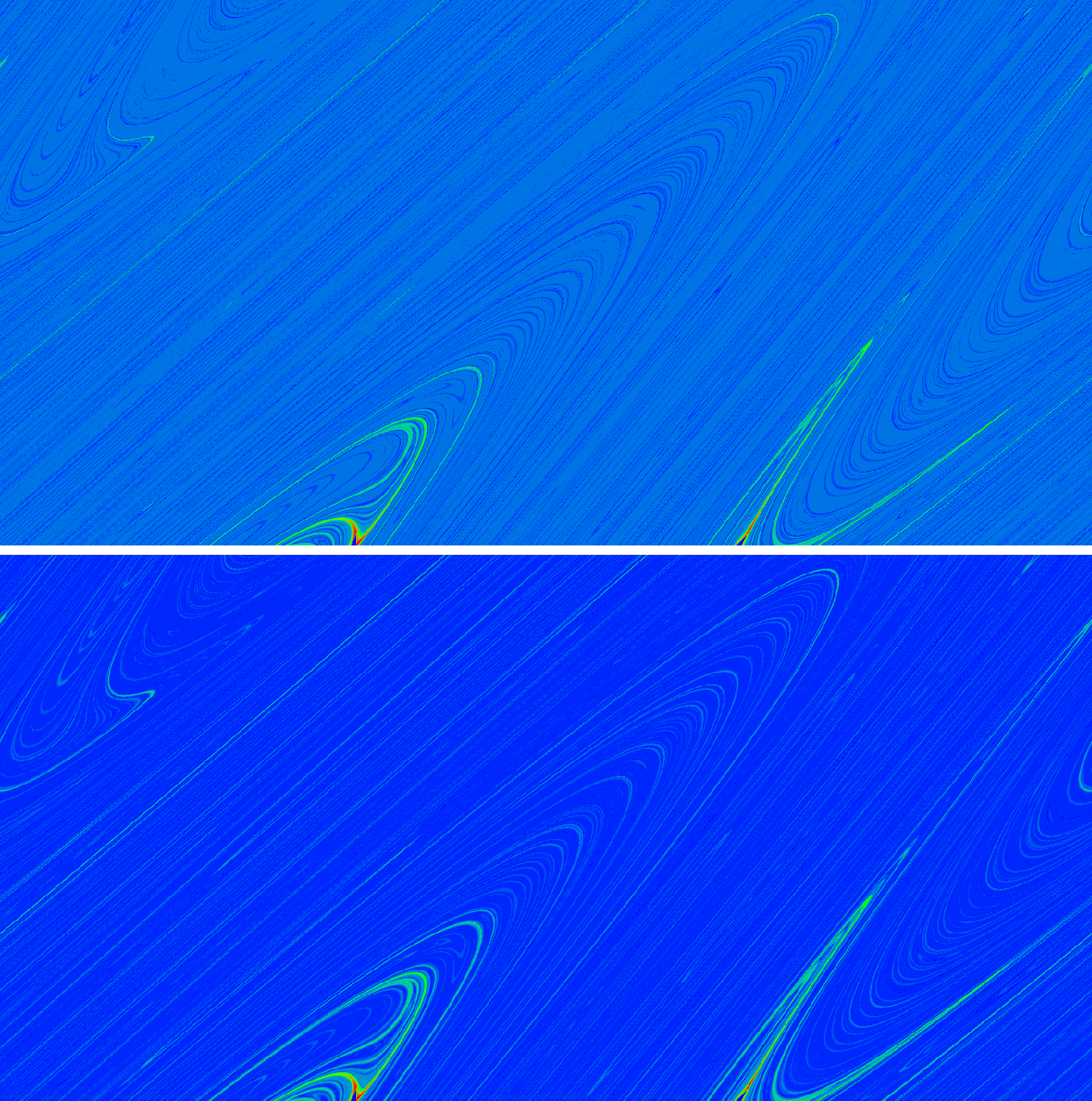}
\end{center}
\caption{\label{fig9} (Color online)
Density plot of the modulus of the components of the 
eigenvectors for $\lambda_1=0.94665516$ (top panel) and 
$\lambda_2=-0.49451923+i\,0.80258270$ (bottom panel) of the UPFO for the 
map (\ref{eq_stmap}) at $K=7$, $M=1600$, $N_d=1279875$ 
and Arnoldi dimension $n_A=1500$.
}
\end{figure}

In view of the exponential distribution of Poincar\'e recurrence times 
as found in  \cite{chsh,chirikov2000b}, a very important question concerns the 
limit of the first non-zero decay rate $\gamma_1(M)$ as $M\to\infty$. 
For the case of critical $K_g$ with a power law distribution we found a 
vanishing limit of $\gamma_1(M)$ but here we expect a finite limit. This is 
indeed the case as can be seen in the right panel of Fig. \ref{fig8} where 
we show $\gamma_1(M)$ as a function of $M^{-1}$. A simple fit with two 
parameters $\gamma_1(M)=d+e\,M^{-1}$ for $M^{-1}\le 0.004$ results in 
$d=0.0994\pm 0.0018$ and $e=20.3$ suggesting the {\em finite} extrapolation 
limit $\lim_{M\to\infty} \gamma_1(M)=0.0994$. However, as can be seen 
in the figure, the quality of the fit is not very good and can be improved by 
a more suitable three parameter fit: 
$\gamma_1(M)=a\,(1+b\,M^{-1})^c$ 
for $M^{-1}\le 0.01$ resulting in $a=0.0857\pm 0.0036$, $b= 1370$ and 
$c=0.389$ suggesting the extrapolation 
limit $\lim_{M\to\infty} \gamma_1(M)=0.0857$ which actually 
coincides (within the error bound) with the ``decay rate'' 
$\gamma_{\rm cl}=2/23.1=0.0866$ found in 
\cite{chirikov2000b} from the exponential tail of the 
distribution of Poincar\'e recurrences. 
\begin{figure}[h]
\begin{center}
\includegraphics[width=0.48\textwidth]{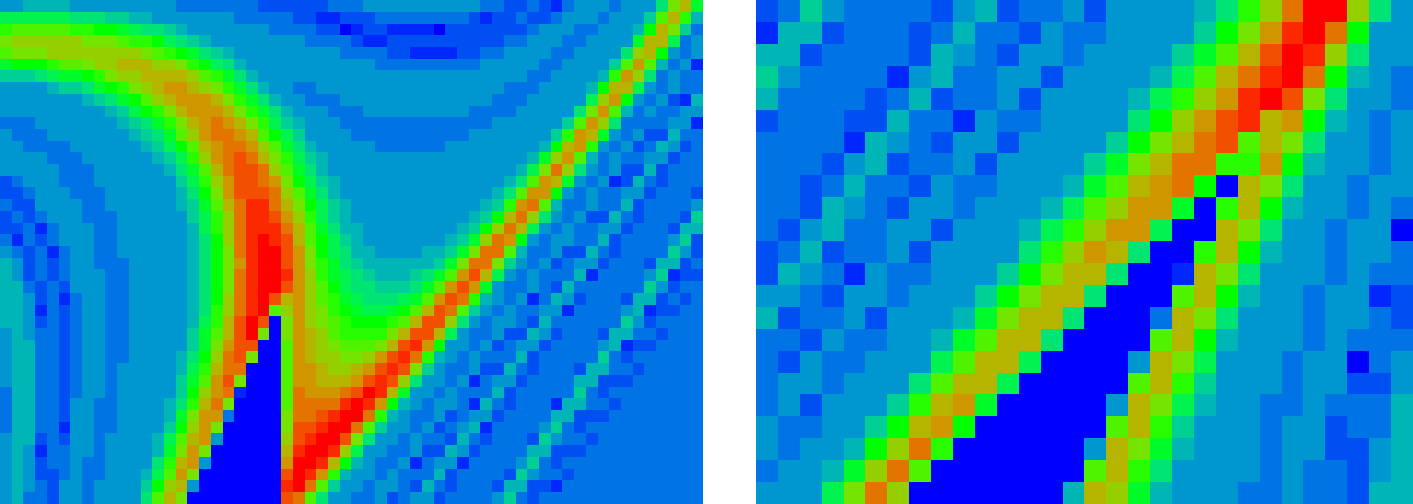}
\end{center}
\caption{\label{fig10} (Color online)
Increased representation of the regions of the two stable islands, 
close to $x=0.33$ and $y=0$ (left panel) or $x=0.67$ and $y=0$ (right 
panel), of the eigenvector for $\lambda_1=0.94665516$ (see top panel 
in Fig. \ref{fig9}). 
}
\end{figure}

Concerning the eigenvectors of $S$, we mention that the eigenvectors 
for the first mode $\lambda_0=1$ represents of course the uniform ergodic 
distribution on the (nearly) full phase space with exception of the 
two small stable islands. 
The eigenvector structure is more interesting for the other (non-uniform) 
modes and in Fig. \ref{fig9} we show for $M=1600$ the density plots of the 
eigenvectors for the two modes $\lambda_1=0.94665516$ and 
$\lambda_2=-0.49451923+i\,0.80258270$. One can clearly identify the 
invariant manifolds and a very interesting structure around the stable 
islands. 

In Fig.~\ref{fig10} we furthermore show zoomed density plots 
of the eigenvector for the 
mode $\lambda_1$ for the two regions close to the stable islands at 
$x=0.33$, $y=0$ and $x=0.67$, $y=0$. Both islands cover 125 out of 
1280000 cells in total with a relative phase 
space volume being approximately
$125/(1600^2/2)\approx 9.7\times10^{-5}$
that, up to statistical fluctuations, 
is in agreement with the result $7.8 \times 10^{-5}$
of \cite{chirikov2000b}.

%%%%%%%%%%%%%%%%%%%%%%%%%%%%%%%%%%%%%%%%%%%%%%%%%%%%%%%%%
\section{Separatrix map at $\Lambda_c=3.1819316$}

In this section, we study the UPFO for a different map, 
called the separatrix map
\cite{chirikov1979}, defined by~:
\begin{equation}
\label{eq_separatrixmap}
{\bar y} = y + \sin (2\pi x) \; , \;\; 
{\bar x} = x + \frac{\Lambda}{2\pi}\ln(|{\bar y}|) \;\; ({\rm mod} \; 1) \;.
\end{equation}
This map can be locally  
approximated by the Chirikov standard map by linearizing the logarithm 
near a certain $y_0$ that leads after rescaling to the map (\ref{eq_stmap})
with an effective parameter 
$K_{\rm eff}=\Lambda/|y_0|$ \cite{chirikov1979}. 
Therefore the separatrix map exhibits strong chaos for small values of 
$|y_0| \ll \Lambda$ while for larger values of $|y_0| $ 
we have the typical KAM-scenario similar to the Chirikov standard map for 
small or modest values of $K$. 

For the separatrix map the width of the chaotic component
 $|y| \leq y_b$ can be estimated from the condition
$K_{\rm eff} \approx \Lambda/y_b \approx 1$ that gives $y_b \approx \Lambda$.
It is known that the golden curve with the rotation number
$r=r_g=(\sqrt{5}+1)/2  = 0.618..$ is critical at
$\Lambda_c=3.1819316$ \cite{chsh} at which there is a quite large chaotic 
domain confined up to values $|y|\le 3.84$. Therefore we define the 
$M\times M$ cells to construct the UPFO
for the phase space domain $0\le x\le 1$, $-4\le y\le 4$.
As in the case of the 
Chirikov standard map, we use a symmetry: $x\to x+1/2\ ({\rm mod}\ 1)$, 
$y\to -y$, to 
reduce this range to $0\le x\le 1$, $0\le y\le 4$ for $M\times M/2$ cells. 
It turns out that for the separatrix map the number of cells visited 
by the trajectory (of length $10^{12}$) 
scales as $N_d \approx C_d M^2/2$ with $C_d \approx 0.78$ 
meaning that the chaotic component contains about $78\%$ of the total area 
of the domain (e.g. $N_d=997045$ for $M=1600$). 
\begin{figure}[h]
\begin{center}
\includegraphics[width=0.48\textwidth]{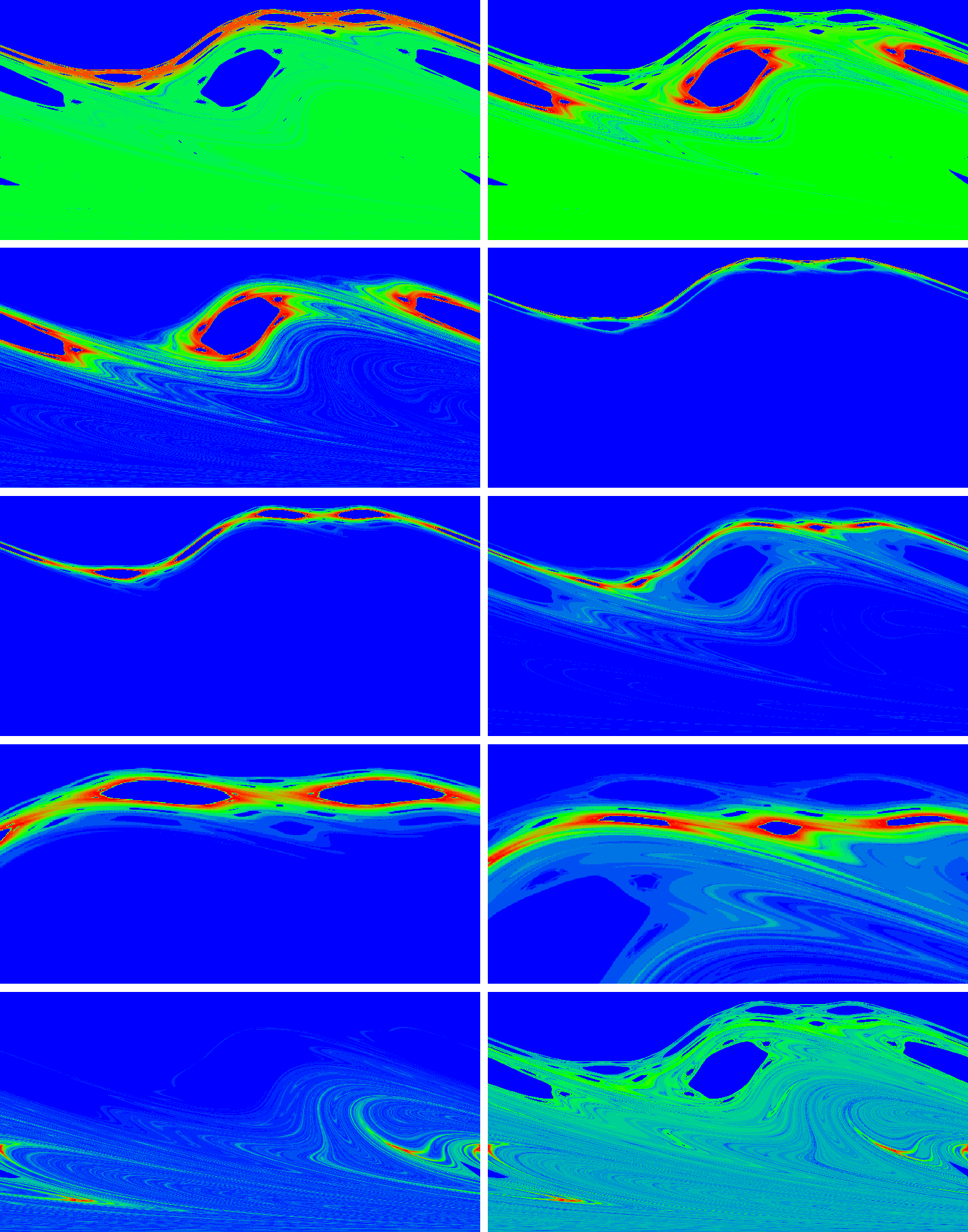}
\end{center}
\caption{\label{fig11} (Color online)
Density plot of the modulus of the components of the 
eigenvectors for $\lambda_1=0.99970603$ (first row, left panel), 
$\lambda_2=0.99828500$ (first row, right panel), 
$\lambda_3=-0.99816880\approx |\lambda_{3}|\,
e^{i\,2\pi(1/2)}$ (second row, left panel), 
$\lambda_{18}=-0.73824747-i\,0.66068553\approx |\lambda_{18}|\,
e^{i\,2\pi(5/13)}$ (second row, right panel),
$\lambda_{20}=-0.80147707+i\,0.58216934\approx |\lambda_{20}|\,
e^{i\,2\pi(2/5)}$ (third row, left panel), and 
$\lambda_{26}=-0.89084450+i\,0.42827996\approx |\lambda_{26}|\,
e^{i\,2\pi(3/7)}$ (third row, right panel)
of the UPFO ($10^{12}$ iterations) for 
the separatrix map at critical $\Lambda_c=3.1819316$, 
$M=1600$, $N_d=997045$ and Arnoldi dimension $n_{\rm A}=3000$. 
In the first three rows and the fifth row the phase space covers the range 
$0\le x\le 1$ and $0\le y\le 4$. 
The fourth row shows zoomed density plots of the eigenvectors for
$\lambda_{20}$ (left panel) and $\lambda_{26}$ (right panel) 
in the phase space range $0.45625\le x\le 0.83125$ and $2.5\le y \le 4$. 
%%%%%%%%%%%%%%%%%%%%%%5
The fifth row shows two modes in the strongly chaotic region for 
$\lambda_{77}=-0.49158867+i\,0.85154001\approx |\lambda_{77}|\,
e^{i\,2\pi(1/3)}$ (left panel) and 
$\lambda_{79}=0.98321618 $ (right panel). 
}
\end{figure}

As in the case of the Chrikiov standard map the matrix $S$ is very sparse 
with small numbers $\kappa\ll N_d$ of non-zero elements per row (or per 
column). The average of these numbers 
(with respect to all rows or all columns) 
is $\kappa_c=\langle \kappa\rangle\approx 19$ (for $M=1600$ and with 
$12< \kappa_c\< 19$ for the other values of $M$). However, 
$\kappa$ has a very large distribution $p(\kappa)$ depending if we 
consider the number of transitions from (or to) a cell which is either 
in the strongly chaotic range for small $y$ or in the range close to the 
critical curve. This distribution has a power law tail 
$p(\kappa)\sim 1/\kappa^2$ for the range
$\kappa_c <\kappa\le \kappa_m$ 
with a very large maximal value $\kappa_m\gg \kappa_c$ (e.g. $\kappa_m=2123$ 
for $M=1600$). We also note that the peak position $\kappa_p$ of 
the distribution $p(\kappa)$ is considerably smaller than 
$\kappa_c$, e.g. $\kappa_p=6$ for $M=1600$. The difference between 
$\kappa_c$ and $\kappa_p$ is clearly due to the long tails of $p(\kappa)$. 
These features of the matrix $S$ are coherent with 
the effective value $K_{\rm eff.}=\Lambda/|y|$ of the chaos parameter
which produces large stretching. This property 
does not create any 
problems for the Arnoldi method and gives only a slight increase of the amount 
of required computer resources (in memory and computational time) since 
the dominant contributions to these resources for 
$1\le n_{\rm A}\ll N_d$ come from terms which do not contain $\kappa_c$ 
(see Section 3). 

We have been able to calculate the full eigenvalue spectrum of $S$ 
for the separatrix map for $25\le M\le 200$ ($279\le N_d\le 16105$) and 
the first 3000 eigenvalues (and the first 500 eigenvectors) 
by the Arnoldi method for $280\le M\le 1600$ ($31273\le N_d\le 997045$).
\begin{figure}[h!!]
\begin{center}
\includegraphics[width=0.48\textwidth]{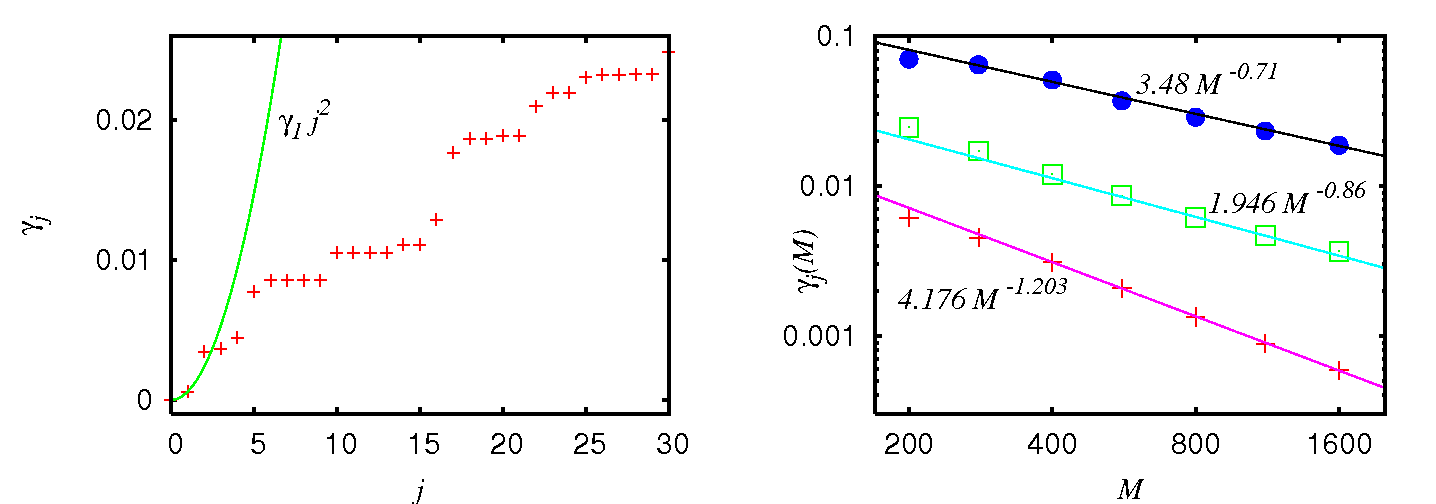}
\end{center}
\caption{\label{fig12} (Color online)
The left panel shows the decay rates $\gamma_j=-2\ln(|\lambda_j|)$ versus 
level number $j$ (red 
crosses) where $\lambda_j$ are the complex eigenvalues of the UPFO 
of the separatrix map for 
$M=1600$ and $N_d=997045$. 
The green curve corresponds to the quadratic dispersion law 
$\gamma_j\approx \gamma_1\,j^2$ which is approximately valid for the 
diffuson modes with $0\le j\le 2$. 
The right panel shows the decay rates $\gamma_j(M)$ for $j=1$ 
(red crosses), $j=3$ (green open squares) and the eigenvector associated 
to the phase $2\pi(5/13)$ (blue full circles, 
see middle right panel in Fig. \ref{fig11}) 
of the UPFO of the separatrix map 
as a function of $M$ in a double logarithmic scale. 
The upper (black) straight line corresponds to the fit 
$\gamma(M)\approx 3.48\,M^{-0.71}$, the middle 
(cyan) line corresponds to the fit 
$\gamma_3(M)\approx 1.946\,M^{-0.86}$ and the lower (magenta) line 
corresponds to the fit $\gamma_1(M)\approx 4.177\,M^{-1.203}$. All fits 
were obtained for the range $400\le M\le 1600$. 
}
\end{figure}

As previously the mode for $\lambda_0$ is uniformly distributed 
in the available phase space and therefore is not shown as a density plot here. 
In Fig. \ref{fig11}, we show the density plot of the more interesting modes 
$\lambda_j$ for $j=1$, $2$, $3$, $18$, $20$, $26$, $77$, $79$ 
(for $M=1600$, $N_d=997045$
and $n_{\rm A}=3000$). The first two of these modes 
(first row in Fig. \ref{fig11}) are similar  to diffuson modes in 
the Chirikov standard map at $K_g$ that is also confirmed 
by the quadratic dispersion of the associated decay rates $\gamma_j$ 
(see left panel of Fig. \ref{fig12}). However, 
the total number of diffuson modes in the list of leading eigenvalues 
is reduced to only three modes (if the uniform mode for $\lambda_0$ is also 
counted as diffuson mode). There are however further diffuson modes 
characterized by real positive eigenvalues $\lambda_j$ close to $1$ 
(e.g. for $j=4,\,5,\,16,\,17$). Actually, if we take out in the 
left panel of Fig. \ref{fig12} the mode for $j=3$ (which corresponds to a 
real negative eigenvalue, see below), the quadratic 
dispersion law extends even up to the first five modes. 

The four modes $\lambda_j$ for $j=3,\,18,\,20,\,26$ 
(second and third rows in Fig. \ref{fig11}) correspond well 
to resonant modes with phases $2\pi(1/2)$, $2\pi(5/13)$, $2\pi(2/5)$, 
$2\pi(3/7)$ and can be identified with the resonances at $1/2$, 
$8/13$, $3/5$ and $4/7$ with $2$, $13$, $5$ and $7$ stable islands
(we remind that phases $2\pi\,\alpha$ and $2\pi\,(1-\alpha)$ are 
always equivalent since they belong to the same pair of complex conjugated 
eigenvalues for $0<\alpha<1/2$). 
For $\lambda_{18}$ (second row, right panel) this is not very clearly visible 
since the resonance $8/13$ is quite small in size as compared to the 
resonance $3/5$ which also contributes to this mode. However, the eigenvector 
components are significantly larger in size at the resonance $8/13$ 
as compared to the resonance $3/5$. 
In the fourth row of Fig. \ref{fig12} we also show zoomed density plots 
of the modes $\lambda_{20}$ (left panel) and $\lambda_{26}$ (right panel) 
(zoom factor $3.2$) in order to visualize clearly the fine structure 
of the resonances. For the mode $\lambda_{20}$ we  see 
(some of) the small islands belonging to the resonance $8/13$ even though this 
mode is more maximal at the resonance $3/5$ 
(the other way round as for $\lambda_{18}$). 
In short these four resonant modes show a similar behavior with phases 
of the form $2\pi(p/q)$ as for the Chirikov standard map at $K_g$. 

We furthermore note that the significant properties of these 6 modes 
(2 diffuson and 4 resonant modes) are essentially determined by the phase 
space region with $y\gtrsim 2$ (KAM region). We have also identified a few 
number of modes which are determined by the strongly chaotic region 
$y\lesssim 2$. In the fifth row of Fig. \ref{fig11} we show 
two of these modes for $\lambda_{77}$ and $\lambda_{79}$. These 
two modes are qualitatively quite similar to the two modes shown in the 
last section for the Chirikov standard map at strong chaos $K=7$ (see 
Fig. \ref{fig9}). This again confirms the picture that the separatrix 
map, at one value of the parameter $\Lambda$, covers implicitly various 
regions with different Chirikov chaos parameters $K_{\rm eff}=\Lambda/|y|$. 
There are also modes which are quite ergodic in the chaotic 
region and those with a 
resonant structure in the KAM region. The higher diffuson 
modes (those with real positive eigenvalues close to $1$) have typically 
a wave node structure in the KAM region, which is quite complicated 
due to the two big islands for the resonance $1/2$, and are simply ergodic 
(or well extended) in the chaotic region. High resolution image files for a 
selected number of these and other modes are available at \cite{qwlib}.

As in the previous sections we have also studied the dependence of some 
of the first non-zero decay rates with $M$ and their scaling behavior for 
$M\to\infty$. As can be seen in the right panel of Fig. \ref{fig12}, 
these decay rates (corresponding to the modes for $\lambda_1$, 
$\lambda_3$ and $\lambda_{18}$ shown in Fig. \ref{fig11}) can be quite well 
fitted (for the values $400\le M \le 1600$) by the power law expressions~: 
$\gamma_1(M)\approx 4.177\,M^{-1.203}$, 
$\gamma_3(M)\approx 1.946\,M^{-0.86}$ and 
$\gamma(M)\approx 3.48\,M^{-0.71}$ where $\gamma(M)$ corresponds to 
the mode $j=18$ for $M=1600$ with phase $2\pi(5/13)$ and localized at 
the resonance $8/13$ 
(however the level number $j$ of this mode changes with $M$ and this 
is not a fit of ``$\gamma_{18}(M)$''). We note that, as for the Chirikov 
standard map at critical $K_g$, these decay rate tend to $0$ for $M\to\infty$. 

%%%%%%%%%%%%%%%%%%%%%%%%%%%%%%%%%%%%%%%%%%%%%%%%%%%%%%%%%
\section{Discussion}

The numerical results for the spectrum and eigenvectors of the UPFO
presented above clearly show that there are modes which
relaxation rates $\gamma \rightarrow 0$ with $M \rightarrow \infty$.
For the map (\ref{eq_stmap}) at $K_g$ we have
$\gamma_1 \sim 1/M^2$ for $M^2 < C^2 = \tau_C \approx 0.66 \times 10^5$
and $\gamma_1 \sim 0.2/M$ for $M^2 > C^2$ (see Fig.~\ref{fig6}).
We interpret this transition in the following way.
According to the results obtained in \cite{chirikov1999}
the average exit time $\tau_n$ from an unstable fixed point of the
Fibonacci approximant $r_n=p_n/q_n$ of the golden rotation number
scales as $\tau_n \approx \tau_g q_n$ with $\tau_g =2.11 \times 10^5$.
Thus even for a moderate value of $q_3=3$ we have a very large exit time 
$\tau_3 \sim 6 \times 10^5$ which is much larger than $1/\gamma_1$
for any $M$ reached numerically. The Ulam method creates effective
noise amplitude $\pm 1/(2M)$ in $x,y$ that generates a diffusion
with the rate $D_U \sim 1/(12 M^2)$. Due to this noise a trajectory
crosses the whole interval $0\leq y \leq 0.38$ up to the golden curve 
on a time scale
$t_U \approx 0.38^2/D_U \approx 1.73 M^2$ which is smaller than $\tau_3$
for $M<600 \sim C$. Thus for $M < C$ the smallest relaxation modes 
have a diffuson type with $\gamma_1 \sim 1/M^2$. For $M \gg C$
we should have $\tau_3 \ll t_U$ and the dominance of the diffuson
modes at low $\gamma$ should disappear.
There is such an indication
in Fig.~\ref{fig6} where  the crossing between $\gamma_1$ and 
$\gamma_6$ should appear at rather large $M$ values.
But at the values $M=1600$ reached in our numerics 
we only start to see an intermediate behavior with
$\gamma_1 \sim 1/M$. Thus we think that the diffuson modes will 
disappear at values $M > 10^4$ which are unfortunately
are out of reach of our numerical data.
Other modes like $\gamma_6$ correspond to sticking of trajectories
in a vicinity of stability islands. However, it is most probable
that the lowest values of $\gamma$ for such modes are also affected 
by the noise of Ulam method for similar reasons as for the diffuson modes
discussed above (but on a smaller scale around main sticking islands).

A similar situation appears also for the separatrix map
where the average exit time $\tau_2 \approx 460$ from a 
vicinity of an unstable fixed
point of the resonance $q=2$ is also rather large
(we determined this time in a similar way as in \cite{chirikov1999}).
This time is significantly smaller than $\tau_3$ of the map (\ref{eq_stmap})
due to strong chaos at $|y|<2$. Due to that we see no $1/M^2$
behavior for $\gamma_1$ and observe only an intermediate behavior
$1/M$ which should disappear at larger values of $M$.
The modes localized around resonant islands are characterized by a
decay of their corresponding lowest $\gamma \sim 1/M^{0.8}$
(see Fig.~\ref{fig12} right panel). This dependence on $M$
clearly shows that these modes are also affected 
by the noise of the Ulam method.
\begin{figure}[h]
\begin{center}
\includegraphics[width=0.48\textwidth]{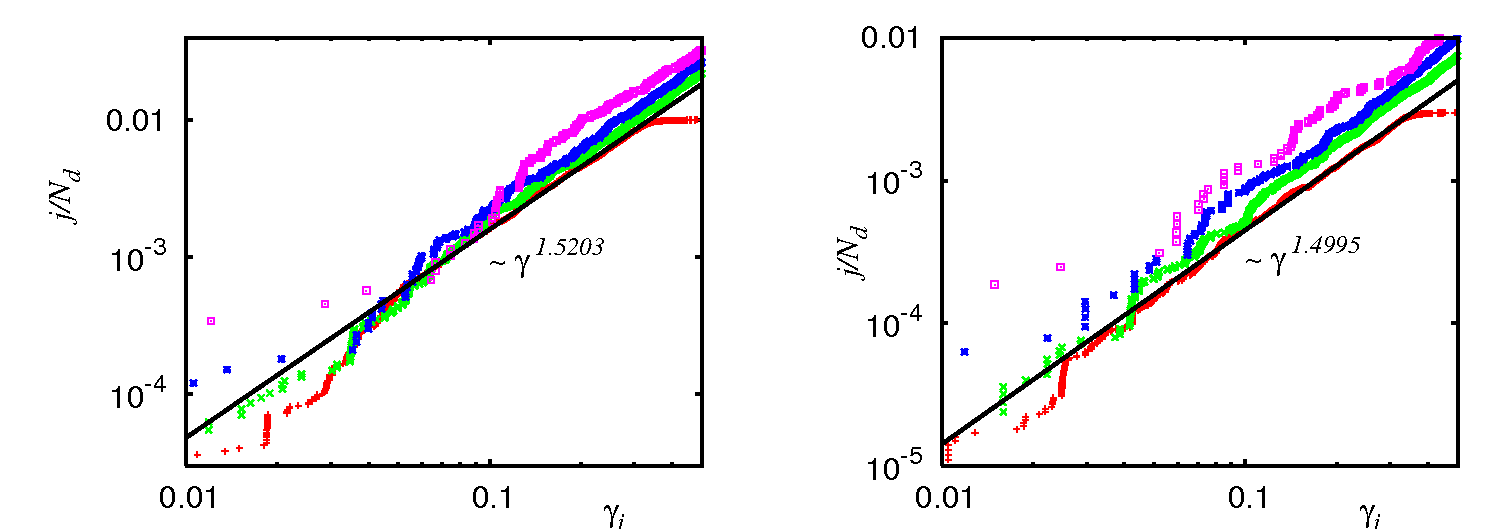}
\end{center}
\caption{\label{fig13} (Color online)
Rescaled level number $j/N_d$ versus the decay rate $\gamma_j$,
in a double logarithmic scale,
for the Chirikov standard map at $K_g$ (left panel) 
and the separatrix map (right panel). 
Red data points correspond to $M=1600$, green
to $M=800$, blue to $M=400$ and magenta to $M=200$
(from bottom to top at $\gamma_j=0.2$). The black
straight line corresponds to the power law fits
$j/N_d\approx 0.052745\,\gamma^{1.5203}$ (left panel) and
$j/N_d\approx 0.014174\,\gamma^{1.4995}$ (right panel) using
the data for $M=1600$ in the range $0.04\le\gamma\le 0.3$.
The statistical error bound of the exponents obtained from the fits
is close to $0.1\%$ in both cases.
Here we used $n_A=3000$ for the Arnoldi method at all $M$,
except the map (\ref{eq_stmap}) at $M=1600$ with $n_A=5000$. 
}
\end{figure}

In fact our aim is to recover the properties of the continuous 
Perron-Frobenius operator using the UPFO as a convergent 
approximant. Our results presented above in Fig.~\ref{fig2}
clearly confirm this
convergence at values $|\lambda|<0.9$. In Fig.~\ref{fig13}
we demonstrate this  convergence even at smaller values of $\gamma$.
Indeed, the data of this figure show that
the integrated density $\rho_\Sigma(\gamma_j)=j/N_d$,
which gives the relative number of states within the interval
$[0,\gamma_j]$, is well described by the dependence
$\rho_\Sigma(\gamma) =A_\Sigma\gamma^{\beta}$
(we remind that we order $\gamma_{j+1} \geq \gamma_{j}$). The prefactor
$A_\Sigma$ varies by a factor $2$ when the matrix size $N_d \propto M^2$
is changed by a factor $64$ (when changing
$M$ from $200$ to $1600$). We attribute this to the fact that
there is a small decrease of effective measure
near the islands with the increase of number of cells.
However, this growth is saturated at large $M$
and we can consider that $A_\Sigma \rightarrow const$ at 
$M \rightarrow \infty$. While a small variation of $A_\Sigma$
with $M$ is visible in Fig.~\ref{fig13} the exponent
$\beta$ remains independent of $N_d$ within few percents
accuracy. For the largest value of $M=1600$ we obtain
$\beta=1.520$ for the Chirikov standard map at $K_g$
and $\beta=1.499$ for the separatrix map
(with a statistical error of 0.1\% by a fit in the range 
$0.04\le\gamma\le 0.3$). Thus our results
show the existence of universal  dependence
$\rho_{\Sigma}(\gamma) \propto \gamma^{1.5}$ independent of $M$.
This dependence works down to smaller and smaller values of $\gamma$
when the size $M$ increases 
(the lowest values of $\gamma_j$  depend on $M$ 
as we discussed above).

Thus our results obtained by the generalized Ulam method
show that the integrated spectral density
decays algebraically at small $\gamma$:
\begin{equation}
\label{eq_gamma}
\rho_{\Sigma}(\gamma) \sim \gamma^\beta \;, \; \beta \approx 1.5\;.
\end{equation}
This behavior leads to an algebraic decay of Poincar\'e
recurrences $P(t) \propto 1/t^\beta$.
Indeed, the probability to stay in a given domain
e.g. $0<y<1/4$ can be estimated as 
$P(t) \sim \int_0^1 (d \rho_{\Sigma}(\gamma)/d \gamma) 
\exp(-\gamma t) d \gamma \sim 1/t^{\beta}$.
The case of $\beta=1/2$ corresponds to a diffusion
on an interval where the diffusion equation 
(\ref{eq_dif}) gives $\gamma_j \sim \pi^2 D_y j^2/(1-r_g)^2$.
In this case $j \propto \sqrt{\gamma_j}$ and
we have $P(t) \sim 1/\sqrt{t}$ 
as discussed in \cite{kiev,chsh} (for $t<1/\gamma_1$).
For $\beta=1.5$ we have the decay 
$P(t) \propto 1/t^{1.5} $ in agreement
with the data for the Poincar\'e recurrences 
found for these two maps (see \cite{chirikov1999,ketzmerick}).

The above arguments 
give an interesting simple relation between the exponent  of 
Poincar\'e recurrences and  the exponent of the spectral
density decay. Of course, our numerical data for the UPFO spectrum
in Fig.~\ref{fig13}  have certain numerical restrictions
showing the algebraic behavior in a moderate range
$0.03 < \gamma < 0.3$. Also it is known that the power law 
decay of $P(t)$ has certain oscillations of the exponent $\beta$.
Thus further studies of the relations between the Poincar\'e
recurrences and the spectrum given by the generalized Ulam method 
are highly desirable.
At the moment, on the basis of our data we make a conjecture
that the both exponents are the same.
These points will be addressed in more detail elsewhere 
\cite{future}. 

In conclusion, our results show that the generalized 
Ulam me\-thod
applied to symplectic maps with divided phase space
converges to the Perron-Frobenius operator of the
continuous map on a chaotic component. The spectrum of this operator
has a power law spectral density of states (\ref{eq_gamma})
for modes with relaxation rates $\gamma \rightarrow 0$.
The exponent of this power law is in agreement with the
exponent of Poincar\'e recurrences decay 
established for such maps, even if 
the range of algebraic decay of the spectral density
is rather moderate compared to the range 
reached for the algebraic decay of the Poincar\'e recurrences. 
More direct relations between
the UPFO and the Poincar\'e recurrences require
further investigations which are in our future plans \cite{future}.

%\newpage
%\vfill
%%%%%%%%%%%%%%%%%%%%%%%%%%%%%%%%%%%%%%%%%%%%%%%%%%%%%%%%%

\end{document}